%% file: impulseControl.tex
\documentclass[aps,prx,twocolumn,english,superscriptaddress,floatfix,longbibliography,amssymb]{revtex4-2}

\usepackage{graphicx}
\usepackage{amsmath}
\usepackage{color}
\usepackage{verbatim}
\usepackage{bm}
\usepackage{cancel}
\usepackage{subfigure}
\usepackage{ulem}

\usepackage{setspace} 

\newcommand{\be}{\begin{equation}}
\newcommand{\ee}{\end{equation}}
\newcommand{\ba}{\begin{eqnarray}}
\newcommand{\ea}{\end{eqnarray}}

\newcommand{\PRA}{Phys. Rev. A}

\newcommand{\PRL}{Phys. Rev. Lett.}
\newcommand{\EPL}{EPL (Europhys. Lett.)}
\newcommand{\RMP}{Rev. Mod. Phys.}

\begin{document}
\singlespacing

\title{Theory of quantum super impulses}

\author{Christopher Jarzynski}

\affiliation{Department of Chemistry and Biochemistry, University of Maryland, College Park, MD 20742 USA}
\affiliation{Institute for Physical Science and Technology, University of Maryland, College Park, MD 20742 USA}
\affiliation{Department of Physics, University of Maryland, College Park, MD 20742 USA}


\begin{abstract}
A quantum impulse is a brief but strong perturbation that produces a sudden change in a wavefunction $\psi({\bf x})$.
We develop a theory of quantum impulses, distinguishing between {\it ordinary} and {\it super} impulses.
An ordinary impulse paints a phase onto $\psi$, while a super impulse -- the main focus of this paper -- deforms the wavefunction under an invertible map, $\mu:{\bf x}\rightarrow{\bf x}^\prime$.
Borrowing tools from optimal mass transport theory and shortcuts to adiabaticity, we show how to design a super impulse that deforms a wavefunction under a desired map $\mu$, and we illustrate our results using solvable examples.
We point out a strong connection between quantum and classical super impulses, expressed in terms of the path integral formulation of quantum mechanics.
We briefly discuss hybrid impulses, in which ordinary and super impulses are applied simultaneously.
While our central results are derived for evolution under the time-dependent Schr\"odinger equation, they apply equally well to the time-dependent Gross-Pitaevskii equation, and thus may be relevant for the manipulation of Bose-Einstein condensates.

\end{abstract}

\pacs{05.70.Ln, etc.}

\maketitle

\section{Introduction}

In introductory classical mechanics, an impulse is a very strong force applied over a very short time, producing a finite momentum change.
Quantum mechanics textbooks do not routinely discuss how such an impulse affects a wavefunction, but the answer is straightforward: the wavefunction acquires a phase,
\be
\label{eq:paintPhase}
\psi_f({\bf x}) = e^{i\Delta S({\bf x})/\hbar} \psi_i({\bf x}) \quad ,
\ee
where $\psi_i$ and $\psi_f$ denote the wavefunction immediately before and after the impulse (see Sec.~\ref{sec:ordinary}).

\begin{figure}[tbp]
\includegraphics[scale=0.4,angle=0]{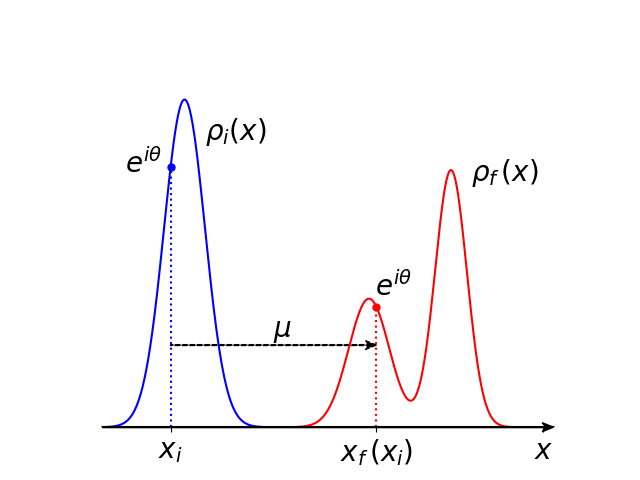}
\caption{
Illustration of a wavefunction deformed under a map $\mu:x_i\rightarrow x_f$, in one dimension.
Writing $\psi_{i,f}$ in the form $\psi = \sqrt{\rho} e^{i\theta}$, the distribution $\rho$ transforms classically under $\mu$ (Eq.~\ref{eq:rho_fi}) and the phase $e^{i\theta}$ carries over (Eq.~\ref{eq:phaseCarriesOver}).
}
\label{fig:deform}
\end{figure}

The present paper concerns {\it super} impulses, whereby a very, {\it very} strong disturbance is applied over a very short time.
The distinction between ordinary and super impulses will be made precise in the next paragraph.
For now we assert that whereas an ordinary impulse paints a phase onto $\psi_i$, as per Eq.~\ref{eq:paintPhase}, a super impulse abruptly {\it deforms} the wavefunction.
This deformation is described by an invertible  map $\mu:{\bf x}_i\rightarrow {\bf x}_f$, through the relation
\be
\label{eq:deform}
\psi_f({\bf x}_f) = \psi_i({\bf x}_i) \left\vert \frac{\partial {\bf x}_f}{\partial {\bf x}_i} \right\vert^{-1/2} \quad , \quad {\bf x}_f = \mu({\bf x}_i) \quad ,
\ee
where $\vert\partial{\bf x}_f/\partial{\bf x}_i\vert$ is the Jacobian of the map $\mu$.
For example, a super impulse might produce a sudden displacement, $\psi_f({\bf x}+{\bf d})=\psi_i({\bf x})$, described by the map $\mu({\bf x}) = {\bf x}+{\bf d}$, or else a sudden stretching or squeezing, $\psi_f(c{\bf x}) = c^{-D/2}\psi_i({\bf x})$, described by $\mu({\bf x}) = c{\bf x}$, where $D$ is the dimensionality of ${\bf x}$-space and $c>0$ is a constant.
More generally, the map need not be linear in ${\bf x}$, as illustrated in Fig.~\ref{fig:deform}.
The aim of this paper is to develop the theory of quantum super impulses, and in particular to show how to design a super impulse that deforms a given initial wavefunction under a desired map.

The general setting we consider involves a system governed by a Hamiltonian
\be
\label{eq:H}
H({\bf x},{\bf p},t) = \frac{{\bf p}^2}{2m} + U_0({\bf x},t) + \frac{1}{\epsilon^k} U_k\left( {\bf x},\frac{t}{\epsilon} \right) \quad,
\ee
where $k \in \{1,2\}$, ${\bf x}\in\mathbb{R}^D$, ${\bf p} = -i\hbar \,\nabla$, $0<\epsilon\ll 1$ and
\be
\label{eq:timeWindow}
U_k({\bf x},\tau) = 0 \quad {\rm for} \quad \tau \notin [0,T] \quad .
\ee
The sum ${\bf p}^2/2m + U_0$ represents a background Hamiltonian, which might describe a particle in a time-dependent potential, or a collection of particles of mass $m$.
In the latter case, ${\bf x}$ includes all particles' coordinates.
The background potential $U_0$ will prove to be irrelevant for our calculations.
The impulse term $U_k/\epsilon^k$ generates a space- and time-dependent force that is applied during an interval from $t=0$ to $\epsilon T$.
We are interested in the system's evolution, under the Schr\" odinger equation, during this interval.
As $\epsilon\rightarrow 0$ with $T$ fixed, the impulse duration becomes infinitesimal and its strength diverges as $\epsilon^{-k}$.
The cases $k=1$ and $2$ define ordinary and super impulses, respectively.
The strength of a super impulse diverges more severely ($\sim\epsilon^{-2}$) than that of an ordinary impulse ($\sim\epsilon^{-1}$), giving rise to the rather different responses described by Eqs.~\ref{eq:paintPhase} and \ref{eq:deform}.

Writing $\psi_{i,f}$ in the form $\psi({\bf x}) = \sqrt{\rho({\bf x})}\exp[i\theta({\bf x})]$, Eq.~\ref{eq:deform} becomes
\begin{subequations}
\label{eq:deform2}
\ba
\label{eq:rho_fi}
\rho_f({\bf x}_f) &=& \rho_i({\bf x}_i) \, \left\vert \frac{\partial {\bf x}_f}{\partial {\bf x}_i} \right\vert^{-1} \\
\label{eq:phaseCarriesOver}
\theta_f({\bf x}_f) &=& \theta_i({\bf x}_i) \quad .
\ea
\end{subequations}
Eq.~\ref{eq:rho_fi} describes the transformation of a probability distribution, when samples drawn from $\rho_i$ are mapped under $\mu$.
We indicate this relation by the notation
\be
\mu:\rho_i \rightarrow \rho_f \quad .
\ee
Under a super impulse, the distribution $\rho = \vert\psi\vert^2$ is transported by the map $\mu$ (Eq.~\ref{eq:rho_fi}), and the phase at ${\bf x}_i$ simply carries over to ${\bf x}_f$ (Eq.~\ref{eq:phaseCarriesOver}), as depicted in Fig.~\ref{fig:deform}.
When referring to the sudden evolution described by Eq.~\ref{eq:deform} or Eq.~\ref{eq:deform2}, we will say that the super impulse ``deforms the wavefunction under the map $\mu$''.

Ordinary and super impulses act on complementary features of a wavefunction.
The former suddenly modifies the wavefunction's phase, without affecting the probability distribution $\vert\psi\vert^2$ (Eq.~\ref{eq:paintPhase}).
The latter suddenly modifies $\vert\psi\vert^2$ (Eq.~\ref{eq:rho_fi}), while leaving the phase unchanged (in the sense of Eq.~\ref{eq:phaseCarriesOver}).

Higher-order impulses, $k\ge 3$, seem to produce divergent behavior as $\epsilon\rightarrow 0$~\footnote{C. Jarzynski, unpublished.}, though this issue deserves to be explored more fully.
Note that an impulse differs from a sudden quench, in which the Hamiltonian instantaneously changes from one operator to another: $H_{t<0} \ne H_{t>0}$.
For an impulse, the ``before'' and ``after'' Hamiltonians are both given by ${\bf p}^2/2m + U_0({\bf x},0)$.

Our motivation for studying quantum super impulses is twofold.
First, Eq.~\ref{eq:deform} represents a novel, asymptotically exact solution of the time-dependent Schr\" odinger equation:
as $\epsilon\rightarrow 0$, the post-impulse wavefunction $\psi_f$ converges to the result given by Eq.~\ref{eq:deform}, without further approximations.
This solution is constructed entirely from classical trajectories (see Sec.~\ref{sec:classSemiclass}), suggesting an interesting equivalence between the impulsive ($\epsilon\rightarrow 0$) and semiclassical ($\hbar\rightarrow 0$) limits.

Second, our paper contributes to a broader effort to develop experimental tools for the rapid manipulation of quantum systems, for instance to protect against environment-induced decoherence~\cite{Zurek03,Schlosshauer19}.
Techniques for accelerating adiabatic evolution known as {\it shortcuts to adiabaticity} (STA)~\cite{Chen10,Guery-Odelin19,Guery-Odelin23} have been studied extensively over the past fifteen years, and have been applied to experimental platforms including cold atoms, trapped ions, superconducting qubits, optical waveguides and diamond NV centers~\cite{Guery-Odelin19}.
Among the various methods in the STA toolkit, the {\it fast-forward} approach~\cite{Masuda08,Masuda10,Torrontegui12,Martinez-Garaot16} accelerates known solutions of the time-dependent Schr\" odinger equation.
Super impulses are extreme shortcuts that (in principle) deform wavefunctions instantaneously, and (in practice) may be useful when sufficiently strong and brief external forces can be applied to a given system.
In Sec.~\ref{sec:design3d} we highlight a link between super impulses and the fast-forward approach, and in Sec.~\ref{sec:discussion} we discuss further connections to STA.

In Sec.~\ref{sec:toy} we introduce a simple system that illustrates the general results we obtain later.
In Sec.~\ref{sec:ordinary} we briefly analyze the evolution of a wavefunction under an ordinary impulse ($k=1$).
In Sec.~\ref{sec:design3d} we show how to construct a quantum super impulse ($k=2$) that achieves the desired deformation, Eq.~\ref{eq:deform}, for a given map $\mu$.
We distinguish between {\it global} and {\it local} super impulses, as explained therein.
Sec.~\ref{sec:examples} illustrates the construction of super impulses with examples for which $U_2$ can be determined analytically.
In Sec.~\ref{sec:classSemiclass} we discuss the close correspondence between quantum super impulses and their classical counterparts.
In Sec.~\ref{sec:hybrid} we briefly discuss {\it hybrid} impulses, in which ordinary and super impulses are applied simultaneously.
We end in Sec.~\ref{sec:discussion} with comments and perspectives.

Throughout this paper, $\mu({\bf x})$ denotes a map, and ${\bf x}_f({\bf x}_i)$ denotes the image of ${\bf x}_i$ under this map.
$\mu(\cdot)$ and ${\bf x}_f(\cdot)$ are identical functions of their arguments.

\section{Toy model}
\label{sec:toy}

Before developing the main ideas of this paper, consider a classical particle in one dimension, and define
\be
\label{eq:toy}
U_1(x,\tau) = -xF_0 \quad,\quad \tau \in [0,T] \quad.
\ee
During an ordinary impulse generated by this potential ($k=1$ in Eq.~\ref{eq:H}), a force $F_0/\epsilon$ is applied over an interval $\epsilon T$.
As $\epsilon\rightarrow 0$, the resulting force $F(t) \rightarrow F_0 T \delta(t)$ produces a sudden change of momentum, with no corresponding change in position:
\be
\label{eq:toy-ordinary}
\Delta x = 0
\quad,\quad
\Delta p = F_0 T \quad.
\ee
This is the textbook case of a classical impulse, typically illustrated by a baseball bat, or a foot, striking a ball.

Now consider a super impulse ($k=2$ in Eq.~\ref{eq:H}), with
\be
\label{eq:U2-toy}
U_2(x,\tau) =
\begin{cases}
-x F_0  \quad &, \quad 0\le\tau<T/2 \\
+x F_0 \quad &, \quad T/2\le\tau\le T
\end{cases}
\ee
Eq.~\ref{eq:U2-toy} describes a uniform force (e.g.\ an electric field acting on a charged particle) that is applied first in one direction, and then in the opposite direction.
Within the interval $t\in[0,\epsilon T]$, the particle undergoes acceleration and then deceleration that scale as $\epsilon^{-2}$, leading to momenta (at intermediate times) that scale as $\epsilon^{-1}$.
In the limit $\epsilon\rightarrow 0$ this super impulse suddenly displaces the particle, with no {\it net} change of momentum:
\be
\label{eq:toy-super}
\Delta x = \frac{F_0 T^2}{4m}
\quad,\quad
\Delta p = 0
\ee

Eqs.~\ref{eq:toy-ordinary} and \ref{eq:toy-super} follow from simple classical calculations.
When a quantum wavefunction is subjected to the same ordinary and super impulses, the resulting sudden changes $\psi_i\rightarrow\psi_f$ are described by, respectively,
\be
\psi_f(x) = e^{i\Delta p x/\hbar} \psi_i(x) \qquad k=1
\ee
with $\Delta p$ given by Eq.~\ref{eq:toy-ordinary}, and
\be
\psi_f(x+\Delta x) = \psi_i(x) \qquad k=2
\ee
with $\Delta x$ given by Eq.~\ref{eq:toy-super}.
These changes closely match the classical ones, and represent simple examples of the more general results obtained in Secs.~\ref{sec:ordinary} and \ref{sec:design3d} below.

Note that the force generated by $U_2$ above satisfies
\be
\label{eq:toy-balance}
\int_0^T F(\tau) \, d\tau = 0 \quad,
\ee
which is a special case of the condition of {\it balance} discussed in Sec.~\ref{sec:classSemiclass}.

\section{Ordinary Impulses}
\label{sec:ordinary}

While super impulses are the primary focus of this paper, let us dispense first with ordinary impulses, during which the wavefunction obeys
\be
\label{eq:se}
i\hbar \frac{\partial\psi}{\partial t} = -\frac{\hbar^2}{2m} \nabla^2\psi + U_0({\bf x},t) \psi + \frac{1}{\epsilon} U_1 \left( {\bf x},\frac{t}{\epsilon} \right) \psi \quad.
\ee
We introduce a convenient {\it fast time} variable,
\be
\tau = \frac{t}{\epsilon} \in [0,T] \quad ,
\ee
which marks the progress of time, from $\tau=0$ to $T$, during the impulse interval $t\in[0,\epsilon T]$.
This variable will be useful in the the analysis of both ordinary and super impulses.
Rewriting Eq.~\ref{eq:se} gives
\be
\label{eq:si-rescaled1}
i\hbar \frac{\partial\psi}{\partial\tau} =  -\epsilon \frac{\hbar^2}{2m} \nabla^2\psi + \epsilon U_0({\bf x},\epsilon\tau) \psi + U_1 ({\bf x},\tau) \psi 
\ee

In the limit $\epsilon\rightarrow 0$, the first two terms on the right vanish and the remaining equation is solved by
\be
\psi({\bf x},\tau) = \exp \left[ -\frac{i}{\hbar} \int_{0}^\tau d\tau^\prime \, U_1({\bf x},\tau^\prime) \right] \psi_i({\bf x}) 
\ee
where $\psi_i({\bf x}) = \psi({\bf x},0)$.
Setting $\tau=T$ we obtain
\be
\label{eq:DeltaS}
\begin{split}
\psi_f({\bf x}) &= e^{i\Delta S({\bf x})/\hbar} \psi_i({\bf x}) \\
\Delta S({\bf x}) &= -\int_{0}^{T} d\tau \, U_1({\bf x},\tau)
\end{split}
\ee
Thus, to paint a phase $\exp[i\Delta S({\bf x})/\hbar]$ onto a wavefunction, we can use an ordinary impulse, setting
\be
\label{eq:design-ordinary-impulse}
U_1({\bf x},\tau) = -\Delta S({\bf x}) \, \nu(\tau)
\ee
where $\int_0^T \nu(\tau)\,d\tau = 1$.
We will use this result in Sec.~\ref{subsec:unsatisfied}.
Note that Eq.~\ref{eq:DeltaS} is valid even if $U_1$ does not factorize as in Eq.~\ref{eq:design-ordinary-impulse}.

Formally, the effect of an ordinary impulse is described by the quantum propagator
\be
\label{eq:K-ordinary}
K({\bf x},\epsilon T \vert {\bf x}_i,0) = e^{i \Delta S({\bf x})/\hbar} \, \delta( {\bf x} - {\bf x}_i ) \quad ,
\ee
where the limit $\epsilon\rightarrow 0$ is understood.

If the impulse potential $U_1$ is independent of $\tau$ (as in Eq.~\ref{eq:toy}), then the last term in Eq.~\ref{eq:H} becomes $U_1({\bf x})T\delta(t)$, and the impulse paints a phase $\exp[-iU_1({\bf x})T/\hbar]$ onto the wavefunction.
This result is well known.
It has been used, for instance, by Ammann and Christensen~\cite{Ammann-PRL97} in their proposed Delta Kick Cooling (DKC) method for cooling atoms.
Eq.~\ref{eq:DeltaS} above is an essentially trivial extension, to $\tau$-dependent $U_1({\bf x},\tau)$, of the known effect of a delta-function ``kick'' on a quantum wavefunction.
DKC and a related earlier approach by Chu {\it et al}~\cite{Chu-OptLett86,Morinaga-PRL99} 
involve the free expansion of an initially trapped particle, or a gas of non-interacting particles, followed by a kick that removes energy excitations.
Dupays and coworkers~\cite{Dupays-PRR21,Dupays-PRA23} have recently extended this approach to scale-invariant dynamics~\cite{Deffner14}, and have allowed for particle-particle interactions.

\section{Super impulses}
\label{sec:design3d}

Now we move on to super impulses.
Since Eq.~\ref{eq:deform} has been asserted but not yet derived, it might seem natural at this point to investigate a wavefunction's evolution under a super impulse, for a given $U_2({\bf x},\tau)$.
It turns out, however, that the evolution $\psi_i\rightarrow\psi_f$ generated by a super impulse diverges as $\epsilon\rightarrow 0$, unless $U_2$ satisfies a {\it balance} condition that generalizes Eq.~\ref{eq:toy-balance}.
We defer a detailed discussion of this condition to Secs.~\ref{sec:classSemiclass} and \ref{sec:discussion}.
Here we instead show how to construct a super impulse that deforms a wavefunction under a chosen map $\mu$.

Our starting point is a wavefunction $\psi_i({\bf x})$, ${\bf x}\in\mathbb{R}^D$, and a continuous, differentiable, invertible map $\mu:{\bf x}_i\rightarrow{\bf x}_f$.
The goal is to design a super impulse that deforms the wavefunction under this map (Eq.~\ref{eq:deform}).

In Sec.~\ref{subsec:satisfied} we assume $\mu$ can be expressed as the gradient of a convex function $\Phi$.
Under this assumption, we show how to construct a super impulse that deforms {\it any} $\psi_i$ under the map $\mu$.
That is, the potential $U_2$ is determined solely by $\mu$ and is independent of $\psi_i$; we will then say that the super impulse is {\it global}.

In Sec.~\ref{subsec:unsatisfied} we show how to proceed when $\mu$ cannot be expressed as the gradient of a convex function.
In that case $U_2$ depends on both $\mu$ and $\psi_i$ (moreover, it must be supplemented by an ordinary impulse, as discussed below), and we will say that the super impulse is {\it local}.

\subsection{$\mu({\bf x})$ is the gradient of a convex function}
\label{subsec:satisfied}

Assume that $\mu:{\bf x}_i\rightarrow{\bf x}_f$ can be written as:
\begin{subequations}
\label{eq:assumption}
\ba
\label{eq:gradient}
{\bf x}_f({\bf x}_i) \equiv \mu({\bf x}_i) &=& \nabla \Phi({\bf x}_i) \\
\label{eq:Phiconvex}
\sum_{j,k=1}^D \frac{\partial^2\Phi}{\partial x_j \partial x_k} \delta x_j \delta x_k &>& 0 \quad \textrm{for all} \quad \delta{\bf x} \ne {\bf 0}
\ea
\end{subequations}
Under this assumption, we first provide a recipe for constructing an impulse potential $U_2({\bf x},\tau)$.
We then verify that, for any initial wavefunction $\psi_i({\bf x})$, the resulting super impulse $\epsilon^{-2}U_2({\bf x},t/\epsilon)$ produces the instantaneous deformation given by Eq.~\ref{eq:deform}, as $\epsilon\rightarrow 0$.
Several technical details are relegated to appendices, as indicated along the way.

To begin, choose a differentiable function $g(\tau)$ satisfying 
\begin{subequations}
\label{eq:g}
\ba
g(0) &=& 0 \quad, \quad g(T) = 1 \\
g(\tau) &\in& (0,1) \quad \textrm{for all} \,\, \tau \in (0,T) \\
\label{eq:gdot}
\dot g(0) &=& \dot g(T) = 0 \quad .
\ea
\end{subequations}
$g(\tau)$ interpolates smoothly, though not necessarily monotonically, from $0$ to $1$.
Here and below, we use the dot ($\dot{~}$) to denote a partial derivative with respect to $\tau$.

Next, combine $\Phi({\bf x})$ and $g(\tau)$ to define
\begin{subequations}
\label{eq:FX-3d}
\ba
\label{eq:Fdef-3d}
F({\bf x}_i,\tau) &=& \frac{1}{2} (1-g) \vert{\bf x}_i\vert^2 + g\Phi({\bf x}_i) 
\\
\label{eq:Xdef-3d}
{\bf X}({\bf x}_i,\tau) &=& \nabla F = (1-g) {\bf x}_i + g {\bf x}_f({\bf x}_i) \quad .
\ea
\end{subequations}
In introducing the functions $F$, ${\bf X}$ and (later) $\Psi$, we follow Aurell {\it et al}~\cite{Aurell-JSP12} (see e.g.\ Eqs.\ 5.8 - 5.10 therein), who used results from optimal mass transport~\cite{Villani03,Peyre19} in their derivation of a refined second law of thermodynamics for overdamped Brownian dynamics.

\begin{figure}[tbp]
\includegraphics[scale=0.4,angle=0]{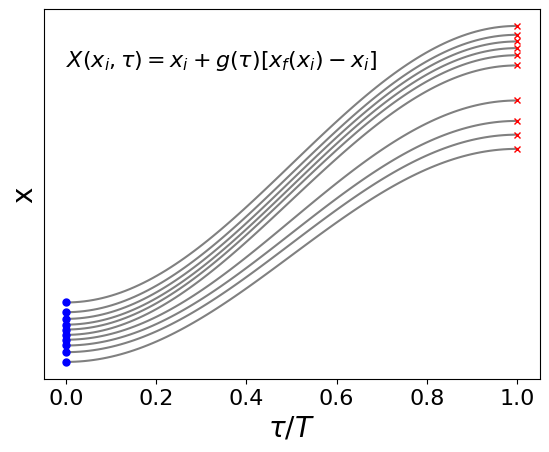}
\caption{
A family of trajectories $x(\tau) = X(x_i,\tau)$ evolving under the velocity field $v(x,\tau)$ (Eq.~\ref{eq:lagr}) from initial conditions $x_i$ (circles) to final conditions $x_f$ (crosses).
The initial conditions shown here are distributed according to $\rho_i(x)$ shown in Fig.~\ref{fig:deform}, and are mapped to final conditions distributed according to $\rho_f(x)$.
As the trajectories do not cross (since $dx_f/dx_i > 0$), the function $X(x_i,\tau)$ can be inverted to give $x_i(X,\tau)$.
Here we chose $g(\tau) = \sin^2(\pi\tau/2T)$, though in general $g(\tau)$ need not increase monotonically.
}
\label{fig:family}
\end{figure}

The function ${\bf X}$ specifies a family of trajectories ${\bf x}(\tau) = {\bf X}({\bf x}_i,\tau)$, parametrized by initial conditions ${\bf x}_i$, that interpolate from ${\bf x}_i$ to ${\bf x}_f({\bf x}_i)$:
\be
\label{eq:xf-3d}
\begin{split}
{\bf x}(0)&={\bf X}({\bf x}_i,0) = {\bf x}_i \\
{\bf x}(T)&={\bf X}({\bf x}_i,T) = {\bf x}_f({\bf x}_i) \quad .
\end{split}
\ee
Fig.~\ref{fig:family} illustrates this construction for a system with one degree of freedom.
When $D=1$, an invertible map $\mu:x_i\rightarrow x_f$ must be either monotonically increasing, $\partial x_f/\partial x_i>0$, or monotonically decreasing, $\partial x_f/\partial x_i<0$.
In both cases the map can be written as the gradient of a function $\Phi(x)$.
By assuming this function is convex (Eq.~\ref{eq:Phiconvex}), we assume $x_f$ increases monotonically with $x_i$, which implies the trajectories $x(\tau)$ do not cross, as shown in Fig.~\ref{fig:family}.

In the general case ($D\ge 1$), since $\Phi({\bf x}_i)$ is convex, so is $F({\bf x}_i,\tau)$, which in turn implies that ${\bf X}({\bf x}_i,\tau)$ is an invertible function of ${\bf x}_i$ (see Appendix \ref{sec:invertibleX}).
We can therefore define a velocity field
\be
\label{eq:vdef-3d}
{\bf v}({\bf X},\tau) = \frac{\partial {\bf X}}{\partial\tau}({\bf x}_i,\tau) = \dot g(\tau) \left[ {\bf x}_f({\bf x}_i) - {\bf x}_i \right] \quad ,
\ee
where ${\bf x}_i={\bf x}_i({\bf X},\tau)$.
By construction, each trajectory ${\bf x}(\tau) = {\bf X}({\bf x}_i,\tau)$ obeys
\be
\label{eq:lagr}
\frac{d{\bf x}}{d\tau} = {\bf v}({\bf x},\tau) \quad,
\ee
and moves along a straight line from ${\bf x}_i$ to ${\bf x}_f({\bf x}_i)$, with a speed proportional to $\dot g(\tau)$, beginning and ending at rest:
\be
\label{eq:v0T3d}
{\bf v}({\bf x},0) = {\bf v}({\bf x},T) = {\bf 0} \quad .
\ee
Borrowing terminology from fluid dynamics, we refer to ${\bf x}(\tau)={\bf X}({\bf x}_i,\tau)$ as a {\it Lagrangian} trajectory, as it evolves under the Lagrangian flow generated by the velocity field ${\bf v}({\bf x},\tau)$.

We similarly construct an acceleration field
\be
\label{eq:adef-3d}
{\bf a}({\bf X},\tau) = \frac{\partial^2{\bf X}}{\partial\tau^2}
= \ddot g(\tau) \left[ {\bf x}_f({\bf x}_i) - {\bf x}_i \right] \quad ,
\ee
which satisfies
\be
\label{eq:av3d}
{\bf a} = ({\bf v}\cdot\nabla) {\bf v} + \dot{\bf v} \quad ,
\ee
as follows by taking the total derivative of both sides of Eq.~\ref{eq:lagr} with respect to $\tau$.
From Eqs.~\ref{eq:gdot} and \ref{eq:adef-3d} it follows that the time-integral of ${\bf a}({\bf X}({\bf x}_i,\tau),\tau)$ over the interval $[0,T]$ vanishes.
Thus if we think of ${\bf F}(\tau) = m{\bf a}$ as the force along this trajectory, then the balance condition
\be
\label{eq:balance}
\int_0^T {\bf F}(\tau) \, d\tau = {\bf 0}
\ee
is satisfied for every Lagrangian trajectory.
This result generalizes the condition imposed in the simple model of Sec.~\ref{sec:toy}, Eq.~\ref{eq:toy-balance}.

Next, we introduce
\be
\label{eq:Psidef}
\begin{split}
\Psi({\bf x},\tau) &= \frac{1}{g(\tau)} \left[
\frac{1}{2} \vert{\bf x}\vert^2 - {\bf x}\cdot{\bf x}_i + F({\bf x}_i,\tau) \right] \\
&= \frac{g}{2} \vert {\bf x}_f - {\bf x}_i \vert^2 - \frac{1}{2} \vert {\bf x}_i \vert^2 + \Phi({\bf x}_i)
\end{split}
\quad ,
\ee
which satisfies the key property (see Appendix~\ref{sec:qsuper3d-app})
\be
\label{eq:nablaPsi}
\nabla\Psi({\bf x},\tau) = {\bf x}_f - {\bf x}_i \quad ,
\ee
where ${\bf x}_i$ and ${\bf x}_f$ are the initial and final conditions of the (unique) Lagrangian trajectory that passes through ${\bf x}$ at time $\tau$.
From Eqs.~\ref{eq:vdef-3d}, \ref{eq:adef-3d} and \ref{eq:nablaPsi} we have
\be
\label{eq:vaPsi-3d}
{\bf v} = \dot g \nabla \Psi \quad,\quad {\bf a} = \ddot g \nabla \Psi \quad .
\ee
These fields play similar roles to velocity and acceleration fields used in Refs.~\cite{Jarzynski17,Patra17} to design fast-forward shortcuts to adiabaticity.
There, however, the aim is to cause a system to evolve non-adiabatically between energy eigenstates (or their classical analogues); in the present work energy eigenstates do not play a privileged role.

Finally, we construct the impulse potential,
\be
\label{eq:UI3d}
U_2({\bf x},\tau) = -m\ddot g \Psi({\bf x},\tau)
\ee
which generates the acceleration and deceleration of each Lagrangian trajectory, through Newton's second law:
\be
{\bf F} = m{\bf a} = -\nabla U_2 \quad .
\ee

We claim that if $U_2({\bf x},\tau)$ is substituted into the Schr\"odinger equation
\be
\label{eq:tdse}
i\hbar \frac{\partial\psi}{\partial t} = H \psi \quad ,
\ee
with
\be
\label{eq:H3d}
H({\bf x},{\bf p},t) = \frac{{\bf p}^2}{2m} + U_0({\bf x},t) + \frac{1}{\epsilon^2} U_2\left( {\bf x},\frac{t}{\epsilon} \right) \quad ,
\ee
and the limit $\epsilon\rightarrow 0$ is taken, then the super impulse at $t=0$ deforms the wavefunction under the map $\mu$.

To establish this result, we first define
\be
\label{eq:Sdef3d}
S({\bf x},\tau) = m\dot g \Psi({\bf x},\tau) \quad ,
\ee
which satisfies
\ba
\label{eq:gradS}
\nabla S &=& m{\bf v}({\bf x},\tau) \\
\label{eq:HJ3d}
\frac{\partial S}{\partial\tau} &+& H_I\left( {\bf x}, \nabla S,\tau \right) = 0 \\
\label{eq:S0T-3d}
S({\bf x},0) &=& S({\bf x},T) = 0 \quad,
\ea
where
\be
\label{eq:HI-3d}
H_I({\bf x},{\bf p},\tau) = \frac{{\bf p}^2}{2m} + U_2({\bf x},\tau) \quad .
\ee
(See Appendix~\ref{sec:qsuper3d-app} for the derivation of Eq.~\ref{eq:HJ3d}.)

For $t\in[0,\epsilon T]$, we now substitute the {\it Ansatz}
\be
\label{eq:ansatz3d}
\psi({\bf x},t) = \sqrt{ \rho({\bf x},\tau) } \,  e^{i\theta({\bf x},\tau)} e^{iS({\bf x},\tau)/\epsilon\hbar}
\ee
with $\tau=t/\epsilon$ into the Schr\"odinger equation, and obtain, as $\epsilon\rightarrow 0$,
\be
\label{eq:theta+continuity3d}
\frac{\partial\theta}{\partial\tau} + {\bf v}\cdot\nabla\theta = 0
\quad,\quad
\frac{\partial\rho}{\partial\tau} + \nabla\cdot\left({\bf v}\rho\right) = 0
\ee
(see Appendix~\ref{sec:qsuper3d-app}).
These equations describe the dynamics under a super impulse, in the fast time variable.

Eq.~\ref{eq:theta+continuity3d}b describes the evolution of the probability density $\rho$ under the flow field ${\bf v}$.
Since this flow maps ${\bf x}_i$ to ${\bf x}_f({\bf x}_i)$ (Eqs.~\ref{eq:xf-3d}, \ref{eq:vdef-3d}), the initial and final densities $\rho_i({\bf x}) = \rho({\bf x},0)$ and $\rho_f({\bf x})=\rho({\bf x},T)$ satisfy
\be
\label{eq:rho_fi-1d}
\rho_f({\bf x}_f) = \rho_i({\bf x}_i) \, \left\vert \frac{\partial {\bf x}_f}{\partial {\bf x}_i} \right\vert^{-1} \quad .
\ee
Eqs.~\ref{eq:vdef-3d} and \ref{eq:theta+continuity3d}a imply $(d/d\tau)\theta({\bf X}({\bf x}_i,\tau),\tau)=0$, hence
\be
\label{eq:theta_fi}
\theta({\bf x}_i,0) = \theta({\bf x}_f,T) \quad .
\ee
Combining these results with Eqs.~\ref{eq:S0T-3d} and \ref{eq:ansatz3d} yields
\be
\label{eq:psif}
\psi_f({\bf x}_f) = \psi_i({\bf x}_i) \left\vert \frac{\partial {\bf x}_f}{\partial {\bf x}_i} \right\vert^{-1/2} \quad ,
\ee
confirming that the super impulse produces the desired deformation of the wavefunction.

Note that the potential $U_2({\bf x},\tau)$ is constructed entirely from the map $\mu({\bf x})$ and interpolating function $g(\tau)$, and does not depend of the choice of $\psi_i({\bf x})$.
Once this potential has been determined, it can be used to deform {\it any} initial $\psi_i$ under the map $\mu$.
To emphasize this feature, we will say that the super impulse generated by $U_2$ is {\it global}.

While the limit $\epsilon\rightarrow 0$ formalizes the notion that an impulse is infinitely brief and strong, in any physical realization the impulse must be finite, hence $\epsilon$ is small but non-zero.
During the impulse, the wavefunction $\psi$ acquires a spatially rapidly varying phase $e^{iS/\epsilon\hbar}$ (Eq.~\ref{eq:ansatz3d}). 
This phase is locally proportional to $e^{i\nabla S\cdot{\bf x}/\epsilon\hbar}$, corresponding to a local momentum ${\bf p} =  \epsilon^{-1}\nabla S$.
As $\vert\psi_i\vert^2$ evolves to $\vert\psi_f\vert^2$, matter is displaced by a finite distance over a time interval $\epsilon T$ (see e.g.\ Fig.~\ref{fig:deform}), implying a momentum that scales as $\epsilon^{-1}$.
Thus the rapidly varying phase in Eq.~\ref{eq:ansatz3d} reflects the rapid displacement of matter that occurs during a super impulse; see also the comments following Eq.~\ref{eq:U2-toy} of Sec.~\ref{sec:toy}.
As $\epsilon\rightarrow 0$ a quantum super impulse produces divergent behavior at intermediate times (Eq.~\ref{eq:ansatz3d}), but the net change in the wavefunction remains finite (Eq.~\ref{eq:deform}).
Analogously, an ordinary classical impulse generates ``infinite'' acceleration at intermediate times, but produces a finite change of momentum.

\subsection{$\mu({\bf x})$ is not the gradient of a convex function}
\label{subsec:unsatisfied}

Now assume that $\mu({\bf x})$ is not the gradient of any convex function, hence the construction of $U_2$ given in Sec.~\ref{subsec:satisfied} breaks down~\footnote{
If we bypass $\Phi$ and $F$, and simply define ${\bf X} = (1-g) {\bf x}_i + g {\bf x}_f$ (see Eq.~\ref{eq:FX-3d}), we have no guarantee that the function ${\bf X}({\bf x}_i,\tau)$ can be inverted to write ${\bf x}_i = {\bf x}_i({\bf X},\tau)$.
}.
In this situation, we do not have a recipe for designing a global super impulse that deforms every $\psi_i$ under the map $\mu$. However, for any particular choice of $\psi_i$, we can design a super impulse, supplemented by an ordinary impulse, as described below, that accomplishes the deformation $\psi_i\rightarrow\psi_f$ given by Eq.~\ref{eq:deform}.
We will say that such a super impulse is {\it local}, to emphasize that it implements the desired deformation for a specific (but arbitrary) choice of $\psi_i$ and not for others.

Given $\psi_i$ and $\mu$, we set $\rho_i = \vert\psi_i\vert^2$ and define $\rho_f$ to be the distribution obtained by transporting $\rho_i$ under $\mu$:
\be
\label{eq:mu-rhoi-rhof}
\mu:\rho_i\rightarrow\rho_f \quad .
\ee
Further define
\be
\label{eq:psif-unsatisfied}
\psi_f({\bf x}) = \sqrt{\rho_f({\bf x})} e^{i\theta_f({\bf x})}
\ee
to be the wavefunction obtained by deforming $\psi_i$ under the map $\mu$.
$\mu$, $\psi_i$, $\psi_f$, $\rho_i$ and $\rho_f$ are taken to be fixed for the remainder of this section.

Now let $\bar\mu:{\bf x}_i\rightarrow\bar{\bf x}_f$ denote an invertible map that satisfies
\ba
\label{eq:mimic}
\bar\mu:\rho_i &\rightarrow& \rho_f \\
\label{eq:gradient-mubar}
\bar{\bf x}_f({\bf x}_i) &\equiv& \bar\mu({\bf x}_i) = \nabla \bar\Phi({\bf x}_i) \quad ,
\ea
for some convex $\bar\Phi({\bf x})$.
That is, both $\mu$ and $\bar\mu$ transport $\rho_i$ to $\rho_f$ (Eqs.~\ref{eq:mu-rhoi-rhof}, \ref{eq:mimic}), and $\bar\mu$ (unlike $\mu$) is the gradient of a convex function (Eq.~\ref{eq:gradient-mubar}).

The existence -- in fact, uniqueness -- of $\bar\mu$ satisfying Eqs.~\ref{eq:mimic}, \ref{eq:gradient-mubar} follows from a theorem due to Brenier~\cite{Brenier-CPAM91,Villani03,Peyre19}:
among all invertible maps $\bar\mu$ that satisfy Eq.~\ref{eq:mimic} (given $\rho_i$ and $\mu$, and with $\rho_f$ determined by Eq.~\ref{eq:mu-rhoi-rhof}), there exists exactly one that is the gradient of a convex function, and this map minimizes the Wasserstein distance
\be
\label{eq:wasserstein}
{\cal W}[\bar\mu] =
\int d^Dx_i \,  \rho_i({\bf x}_i) \, \left\vert \bar{\bf x}_f({\bf x}_i) - {\bf x}_i \right\vert^2 \quad .
\ee
$\bar\mu$ can be constructed numerically using the method of Benamou and Brenier~\cite{Benamou-NM00}.
In what follows, we will make use of the existence of $\bar\mu$ satisfying Eqs.~\ref{eq:mimic} and \ref{eq:gradient-mubar}, but will not need the property that $\bar\mu$ minimizes ${\cal W}$.

We now apply the construction of $U_2({\bf x},\tau)$ given in Eqs.~\ref{eq:g} - \ref{eq:UI3d} to the map $\bar\mu$.
The result is a super impulse that produces the deformation
\be
\psi_i({\bf x}) = \sqrt{\rho_i({\bf x})} e^{i\theta_i({\bf x})} \quad \rightarrow \quad \sqrt{\bar\rho_f({\bf x})} e^{i\bar\theta_f({\bf x})} = \bar\psi_f({\bf x})
\ee
with $\bar\rho_f = \rho_f$ (by Eq.~\ref{eq:mimic}), but in general
\be
\bar\theta_f({\bf x}) \ne  \theta_f({\bf x})
\ee
(see Eq.~\ref{eq:psif-unsatisfied}), since $\bar\mu\ne\mu$.
Thus the super impulse constructed from $\bar\mu$ produces a wavefunction $\bar\psi_f$ with the correct final amplitude but the wrong phase.
This phase error can be corrected by acting on $\bar\psi_f$ with an ordinary impulse, obtained by setting
\be
\Delta S({\bf x}) = \hbar \left[ \theta_f({\bf x}) - \bar\theta_f({\bf x}) \right]
\ee
in Eq.~\ref{eq:design-ordinary-impulse}.
This ordinary impulse erases the wrong phase and paints the right one.

We conclude that for a particular choice of $\psi_i$, a super impulse constructed from the map $\bar\mu$, supplemented by an appropriately tailored ordinary impulse, produces the desired final wavefunction $\psi_f$ satisfying Eq.~\ref{eq:deform}.
As mentioned above, this recipe is local: different choices of $\psi_i$ lead to different impulse potentials $U_2$.

\begin{figure}[tbp]
\includegraphics[scale=0.4,angle=0]{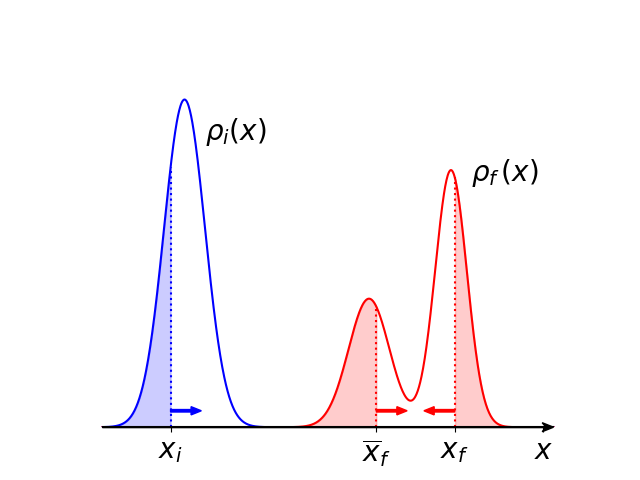}
\caption{
Given a decreasing map $\mu:x_i\rightarrow x_f$ that transports $\rho_i$ to $\rho_f$, the increasing map $\bar\mu:x_i\rightarrow \bar x_f$ defined by Eq.~\ref{eq:mubardef} also transports $\rho_i$ to $\rho_f$.
The shaded regions have equal areas, given by $c_i(x_i)$, $c_f(\bar x_f)$ and $1-c_f(x_f)$.
The stubby arrows indicate that $\bar x_f$ increases, and $x_f$ decreases, with $x_i$.
}
\label{fig:mumubar}
\end{figure}

The map $\bar\mu$ satisfying Eqs.~\ref{eq:mimic} and \ref{eq:gradient-mubar} is particularly easily determined when $D=1$.
In one dimension, there is a unique monotonically increasing map that transports $\rho_i$ to $\rho_f$, and a unique monotonically decreasing map that does the same.
These maps are determined using cumulative distribution functions, $c(x) = \int_{-\infty}^x dx^\prime \, \rho(x^\prime)$, as depicted in Fig.~\ref{fig:mumubar}.
If $\mu$ is a monotonically decreasing map that transports $\rho_i$ to $\rho_f$ (hence it cannot be written as the gradient of a convex function), then the monotonically increasing map $\bar\mu:x_i\rightarrow\bar x_f$ that transports $\rho_i$ to $\rho_f$ is given by
\be
\label{eq:mubardef}
\bar x_f(x_i) = c_f^{-1}(c_i(x_i)) \quad .
\ee

\section{Examples}
\label{sec:examples}

Here we illustrate the methods for constructing the impulse potential $U_2$ described in Section \ref{sec:design3d}, using four maps for which $U_2$ can be determined analytically.
These examples cover both global and local maps, and both the $D=1$ and $D>1$ cases.
Throughout this section, the interpolating function $g(\tau)$ satisfies Eq.~\ref{eq:g} but is otherwise unspecified.

\subsubsection{Global super impulse in one degree of freedom}
\label{subsec:splitting}

\begin{figure}[tbp]
   \subfigure[]{
   \label{fig:global-1d-X_xi}
   \includegraphics[scale=0.35,angle=0]{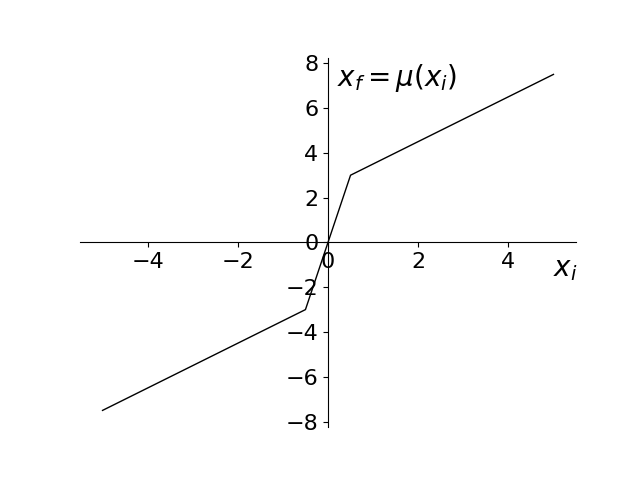}
   }
   \subfigure[]{
   \label{fig:global-1d-deform}
   \includegraphics[scale=0.35,angle=0]{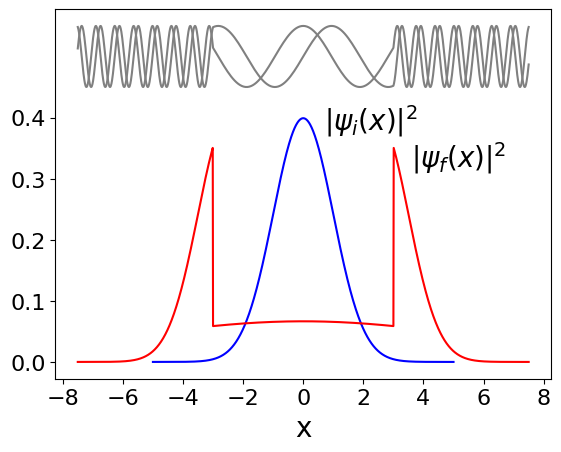}
   }
   \subfigure[]{ 
   \label{fig:global-1d-trajs}
   \includegraphics[scale=0.35,angle=0]{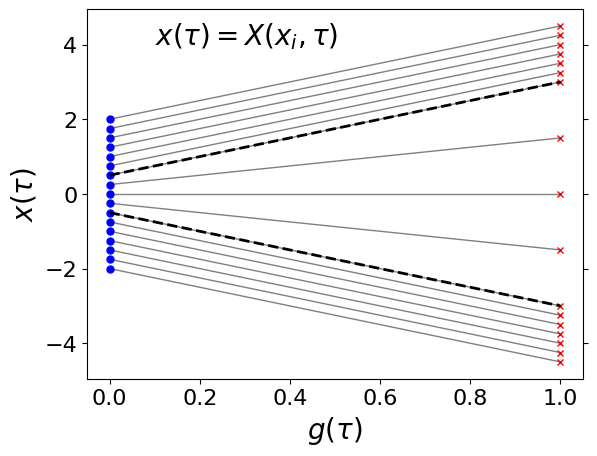}
   } 
\caption{(a) The map $\mu$ defined by Eq.~\ref{eq:mu-1d-global}, with $a=0.5$, $b=3.0$.
(b) $\mu$ deforms a Gaussian wavepacket $\psi_i(x)$ by cleaving it into left and right portions. The offset oscillatory curves show the real and imaginary parts of the phase of $\psi_f(x)$.
(c) Trajectories ${\bf x}(\tau)$ plotted against $g(\tau)$.  Dashed lines indicate the trajectories from $x_i=\pm a$ to $x_f=\pm b$.
}
\label{fig:global-1d}
\end{figure}

Consider the map (with $b>a>0$)
\be
\label{eq:mu-1d-global}
\mu: x_i \rightarrow x_f = 
\begin{cases}
x_i - b + a & \text{if $x_i <  -a$} \\
(b/a) x_i & \text{if $-a \le x_i \le a$} \\
x_i -a + b & \text{if $x_i >  a$} 
\end{cases} \quad ,
\ee
shown in Fig.~\ref{fig:global-1d-X_xi} for $a=0.5$, $b=3.0$.
Under this map, an initial wavefunction $\psi_i(x)$ is cleaved: the portion corresponding to $x<-a$ is shifted leftward by $b-a$, the portion corresponding to $x>a$ is shifted rightward by the same distance, and the in-between portion is stretched linearly.
Specifically, Eqs.~\ref{eq:deform} and \ref{eq:mu-1d-global} give
\be
\label{eq:psi-1d-global}
\psi_f(x) =
\begin{cases}
\psi_i(x+b-a) & \quad,\quad x < -b \\
\sqrt{a/b} \,\, \psi_i(ax/b) & \quad,\quad -b \le x \le b \\
\psi_i(x-b+a) & \quad,\quad x > b
\end{cases}
\quad ,
\ee
as illustrated in Fig.~\ref{fig:global-1d-deform} for the Gaussian wavepacket 
\be
\label{eq:Gnpacket}
\psi_i(x) = \left(2\pi\sigma^2\right)^{-1/4} e^{-x^2/4\sigma^2} e^{ikx} \quad,
\ee
with $\sigma=1$, $k=10$.

Since $\mu(x)$ increases monotonically with $x$, we can construct a global impulse potential $U_2(x,\tau)$ that deforms any $\psi_i(x)$ according to Eq.~\ref{eq:psi-1d-global}.

The map $\mu$ produces Lagrangian trajectories $x(\tau)$ illustrated in Fig.~\ref{fig:global-1d-trajs}.
Trajectories with initial conditions $\vert x_i\vert > a$ move leftward or rightward, at speed $\dot g (b-a)$, while the trajectories in between ($\vert x_i\vert\le a$) spread out from one another, as the interval $[-a,+a]$ is stretched into the interval $[-b,+b]$.
Following the steps described in Sec.~\ref{subsec:satisfied}, we obtain, after some algebra,
\be
\label{eq:UI-global-1d}
U_2(x,\tau) = -m\ddot g (b-a) \cdot
\begin{cases}
\vert x \vert-(c/2) & \vert x\vert >c \\
x^2/2c & \vert x\vert \le c
\end{cases}
\ee
where
\be
c(\tau) = (1-g)a + g b = X(a,\tau) \quad .
\ee
The condition $\vert x(\tau)\vert > c(\tau)$ is equivalent to $\vert x_i\vert > a$.

When $\ddot g(\tau) > 0$, the potential $U_2(x,\tau)$ forms an inverted ``V'', whose vertex is rounded in the region $\vert x\vert \le c$.
This rounded, quadratic region of the potential governs the trajectories located between the two dashed lines in Fig.~\ref{fig:global-1d-trajs}, causing them to move away from one another.
The leftward- and rightward-moving trajectories outside this region (parallel lines in Fig.~\ref{fig:global-1d-trajs}) evolve under the linear portions of this V-shaped potential, where $\vert x\vert >c$.

The impulse potential $U_2$ is discontinuous in its second derivative at $x=\pm c$, resulting in discontinuities in $\psi_f(x)$ at $x=\pm b$, see Eq.~\ref{eq:psi-1d-global} and Fig.~\ref{fig:global-1d-deform}.
The map
\be
\mu:x_i \rightarrow x_i + (b-a) \tanh \left( \frac{x_i}{a} \right)
\ee
provides a smoothed version of the one defined by Eq.~\ref{eq:mu-1d-global}, leading to a potential $U_2$ and softly cleaved final wavefunction $\psi_f$ that are continuous in all derivatives.
For this map we have $X(x_i,\tau) = x_i + g(\tau)(b-a)\tanh(x_i/a)$, which cannot be inverted analytically.
However, we can readily construct $x_i(X,\tau)$ numerically, leading to a numerical solution for $U_2(x,\tau)$.

\subsubsection{Local super impulse in one degree of freedom}

Now consider the reflection map
\be
\label{eq:mu-1d-local}
\mu:x_i\rightarrow -x_i \quad .
\ee
Since $\mu(x)$ decreases monotonically with $x$, we cannot construct a global super impulse for this map by following the recipe described in Sec.~\ref{subsec:satisfied}.
Instead, we illustrate the construction of a local super impulse, which deforms a specific choice of $\psi_i$ under the map $\mu$.
As described in Sec.~\ref{subsec:unsatisfied}, this super impulse must be supplemented by an ordinary impulse that ``corrects'' the phase of $\psi_f$.

We choose
\be
\label{eq:psii-example-1d}
\psi_i(x) = \left(2\pi\sigma^2\right)^{-1/4} e^{-(x-s)^2/4\sigma^2} e^{ikx} \quad,\quad s > 0.
\ee
Under the map $\mu$, this Gaussian wavepacket is reflected around the origin:
\be
\label{eq:psif-example-1d}
\psi_f(x) = \left(2\pi\sigma^2\right)^{-1/4} e^{-(x+s)^2/4\sigma^2} e^{-ikx} \quad.
\ee
We wish to construct a super impulse, supplemented by an ordinary impulse, that generates this deformation, for this particular choice of $\psi_i$.
We follow the steps described in Sec.~\ref{subsec:unsatisfied}.

The cumulative distribution functions $c_{i,f}$ obtained from $\psi_{i,f}$ satisfy $c_f(x-2s) = c_i(x)$, hence Eq.~\ref{eq:mubardef} gives
\be
\bar\mu:x_i\rightarrow x_i-2s \quad .
\ee
Applying Eqs.~\ref{eq:g} - \ref{eq:UI3d} to the map $\bar\mu$, we get $X=x_i - 2sg$, $v=-2s\dot g$, and
\be
\label{eq:UI-local-1d}
U_2(x,\tau) = 2msx\ddot g(\tau) \quad .
\ee
This potential generates a force $-2ms\ddot g$, and the resulting super impulse simply displaces $\psi_i$ by $-2s$:
\be
\psi_i(x) \rightarrow
\bar\psi_f(x) = \left(2\pi\sigma^2\right)^{-1/4} e^{-(x+s)^2/4\sigma^2} e^{ikx} \quad.
\ee
We then use an ordinary impulse (Sec.~\ref{sec:ordinary}) to paint a phase $e^{-2ikx}$ onto $\bar\psi_f$, arriving at the desired $\psi_f$ (Eq.~\ref{eq:psif-example-1d}).

The super impulse generated by Eq.~\ref{eq:UI-local-1d} displaces {\it any} $\psi_i$ by $-2s$.
Because the distribution $\rho_i = \vert\psi_i\vert^2$ given by Eq.~\ref{eq:psii-example-1d} happens to be symmetric about the point $x=s$, the displacement of $\psi_i$ by $-2s$ is equivalent, apart from a phase, to reflection about the point $x=0$ (Eq.~\ref{eq:mu-1d-local}).
For a generic choice of $\psi_i$, the steps described in Sec.~\ref{subsec:unsatisfied} lead to a potential $U_2$ that is not linear in $x$, and the resulting local super impulse does not simply displace the wavefunction.

We used the reflection map, Eq.~\ref{eq:mu-1d-local}, to illustrate the approach of Sec.~\ref{subsec:unsatisfied}.
However, it is a peculiarity of this map that it 
{\it can} be implemented -- up to an overall phase -- with a global super impulse, using the potential
\be
\label{eq:U2-ho}
U_2(x,\tau) = \frac{1}{2} m\omega^2 x^2 \quad,\quad \omega = \frac{\pi}{T} \quad .
\ee
During this super impulse, the wavefunction evolves under the Schr\"odinger equation for a harmonic oscillator, for exactly half a period of oscillation.
This evolution produces a final wavefunction $\psi_f(x) = -i \psi_i(-x)$
\footnote{
This result is easily derived by decomposing $\psi_i$ into harmonic oscillator energy eigenstates, and using the familiar expression for the energy eigenvalues to evolve each term in the decomposition.}.
We stress that Eq.~\ref{eq:U2-ho} does {\it not} follow by applying the recipe of Sec.~\ref{subsec:satisfied} to the reflection map, and should be viewed as a special solution that is specific to this map.

\subsubsection{Global super impulse in $D>1$ degrees of freedom}
\label{subsec:global-3d}

Consider the linear map
\be
\label{eq:mu-3d-global}
\mu:{\bf x}_i \rightarrow M{\bf x}_i
\ee
where the $D\times D$ real matrix $M$ is symmetric and positive definite.
These conditions guarantee that $\mu({\bf x})$ is the gradient of a convex function:
\be
\label{eq:mugrad}
\mu({\bf x}) = M{\bf x} = \nabla \left( \frac{1}{2} {\bf x}^T M {\bf x} \right) \quad.
\ee
The map $\mu$ performs linear rescaling along the eigenvectors of $M$.
Letting $\hat e_j$ and $\lambda_j$ denote these eigenvectors and associated eigenvalues, the map acts as follows:
\be
{\bf x}_i = \sum_{j=1}^D \chi_j \hat e_j  \quad \rightarrow \quad 
{\bf x}_f = \sum_{j=1}^D \lambda_j\chi_j \hat e_j \quad .
\ee
with $\hat e_i\cdot\hat e_j=\delta_{ij}$ and all $\lambda_j>0$.
This map stretches and/or contracts the coordinates along the orthogonal directions $\{ \hat e_j \}$.

Following Sec.~\ref{subsec:satisfied} and defining $B(\tau) = (1-g) I + g  M$, we obtain
\ba
{\bf X}({\bf x}_i,\tau) &=& B {\bf x}_i  \\
\label{eq:UI-global-3d}
U_2({\bf x},\tau) &=& -\frac{1}{2} m\ddot g \, {\bf x}^T ( M - I) B^{-1} {\bf x} \quad .
\ea
The super impulse generated by Eq.~\ref{eq:UI-global-3d} is global: it deforms any $\psi_i$ under the map $\mu$ given by Eq.~\ref{eq:mu-3d-global}.

When $D=1$, $M$ becomes a positive constant, $U_2$ is a time-dependent quadratic potential, and the super impulse linearly stretches or contracts the wavefunction.

\subsubsection{Local super impulse in $D>1$ degrees of freedom}

Finally, taking $D=3$, consider the map
\be
\label{eq:map-local3d}
\mu:{\bf x}_i \rightarrow M{\bf x}_i \quad,\quad M = 
\left(
\begin{array}{rrr}
a & -a & ~0 \\
a & a & ~0 \\
0 & 0 & ~1
\end{array}
\right)
\quad,\quad
a = \sqrt{\frac{1}{2}} ,
\ee
which performs a $\pi/4$ rotation around the $z$ axis.
Because $M$ is not symmetric, the function $\mu({\bf x}) = M{\bf x}$ is not the gradient of a scalar function $\Phi({\bf x})$~\footnote{
E.g.\ Eq.~\ref{eq:map-local3d} implies $\partial x_f/\partial y_i \ne \partial y_f/\partial x_i$, which is inconsistent with ${\bf x}_f = \nabla\Phi({\bf x}_i)$, where ${\bf x}=(x,y,z)$.
}.
We illustrate how to implement $\mu$ locally, for a particular choice of $\psi_i$.

\begin{figure}[tbp]
\includegraphics[scale=0.5,angle=0]{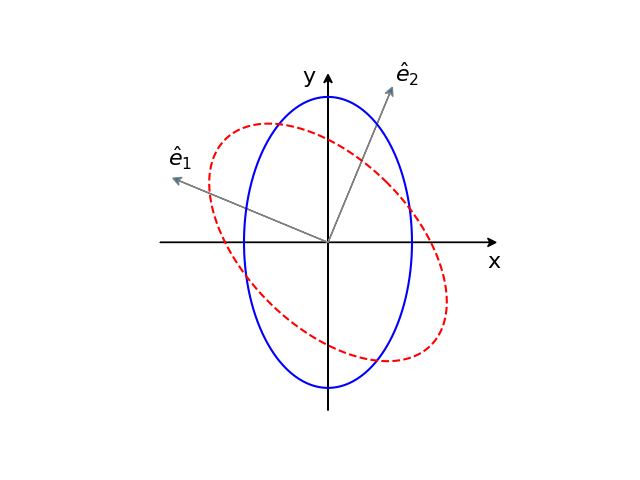}
\caption{
{\it Solid blue curve}: contour of $\rho_i({\bf x}) = \vert\psi_i({\bf x})\vert^2$ (Eqs.~\ref{eq:psi_i-local3d}, \ref{eq:rhoi-example-3d}) at $z=0$.
{\it Dashed red curve}: corresponding contour of $\rho_f({\bf x})$, the image of $\rho_i({\bf x})$ under the map $\mu$, which performs a $\pi/4$ rotation around the $z$ axis (Eqs.~\ref{eq:map-local3d}, \ref{eq:E}).
$\rho_f({\bf x})$ is identical to $\bar\rho_f({\bf x})$, the image of $\rho_i({\bf x})$ under the map $\bar\mu$ (Eqs.~\ref{eq:mubar-local3d}, \ref{eq:Ebar}), which performs a linear stretch and contraction, respectively, along the eigenvectors of $\bar M$, $\hat e_{1,2}\propto (-a,1\mp a,0)$, with eigenvalues $\lambda_{1,2} = (4\pm\sqrt{2})b$, where $a=\sqrt{1/2}$ and $b=\sqrt{1/14}$.
}
\label{fig:local-3d}
\end{figure}

We choose
\be
\label{eq:psi_i-local3d}
\psi_i({\bf x})
= \sqrt{{\cal N}} \, e^{-(3/4)x^2 -(1/4)y^2 -(3/4)z^2}
\ee
with ${\cal N} = (9/8\pi^3)^{1/2}$.
The distribution 
\be
\label{eq:rhoi-example-3d}
\rho_i({\bf x}) = \vert\psi_i\vert^2 = {\cal N} e^{-{\bf x}^T D {\bf x}/2 }
\quad,\quad
D =
\left(
\begin{array}{rrr}
3 & 0 & 0 \\
0 & 1 & 0 \\
0 & 0 & 3
\end{array}
\right)
\ee
is a Gaussian distribution whose contours are cigar-shaped ellipsoids oriented along the $y$-axis.
Under the map $\mu$, this distribution is rotated by $\pi/4$ around the $z$-axis, as illustrated in Fig.~\ref{fig:local-3d}.
Using $\rho_f(M{\bf x}) = \rho_i({\bf x})$ (since $\det M=1$) we have
\be
\label{eq:E}
\rho_f({\bf x}) = {\cal N} e^{-{\bf x}^T E {\bf x}/2 }
\quad,\quad
M^T E M = D
\ee

Now consider the map
\be
\label{eq:mubar-local3d}
\bar\mu:{\bf x}_i \rightarrow \bar M{\bf x}_i \quad,\quad \bar M = 
\left(
\begin{array}{rrr}
5b & -b & ~0 \\
-b & 3b & ~0 \\
0 & 0 & ~1
\end{array}
\right)
\quad,\quad
b = \sqrt{\frac{1}{14}} .
\ee
Note that $\det\bar M = 1$.
This map performs a linear stretch by a factor $\lambda_1$ along the eigenvector $\hat e_1$ of $\bar M$,
and a linear contraction by a factor $\lambda_2$ along $\hat e_2$; see Fig.~\ref{fig:local-3d}.
Under this map, $\rho_i$ given by Eq.~\ref{eq:rhoi-example-3d} transforms to $\bar\rho_f$ that satisfies $\bar\rho_f(\bar M {\bf x}) = \rho_i({\bf x})$, hence
\be
\label{eq:Ebar}
\bar \rho_f({\bf x}) = {\cal N} e^{-{\bf x}^T \bar E {\bf x}/2 }
\quad,\quad
\bar M^T \bar E \bar M = D \quad .
\ee

The matrices $E$ and $\bar E$ define Gaussian distributions $\rho_f$ and $\bar\rho_f$.
From Eqs.~\ref{eq:E} and \ref{eq:Ebar} we obtain, by direct evaluation,
\be
E = \bar E = 
\left(
\begin{array}{rrr}
2 & 1 & 0 \\
1 & 2 & 0 \\
0 & 0 & 1
\end{array}
\right)
\quad,
\ee
therefore $\rho_f=\bar\rho_f$.
Thus, while the maps $\mu$ and $\bar\mu$ differ -- the former rotates around the $z$-axis, the latter stretches and contracts in the $xy$-plane -- their effects on the particular distribution $\rho_i$ (Eq.~\ref{eq:rhoi-example-3d}) are the same.

Since $\bar M$ is symmetric and positive definite, $\bar\mu({\bf x})$ is the gradient of a convex function (Eq.~\ref{eq:mugrad}).
Proceeding as in Sec.~\ref{subsec:global-3d} we obtain
\be
\label{eq:UI-local-3d}
U_2({\bf x},\tau) = -\frac{1}{2} m\ddot g \, {\bf x}^T (\bar M - I) B^{-1} {\bf x} \quad ,
\ee
where
\be
B(\tau) = (1-g) I + g \bar M \quad .
\ee
An expression for $B^{-1}(\tau)$ can be obtained analytically, but is not particularly illuminating.

For the specific choice of $\psi_i$ given by Eq.~\ref{eq:psi_i-local3d}, the potential $U_2$ generates a super impulse whose net effect is a $\pi/4$ rotation around the $z$ axis, thereby implementing the map $\mu$ locally.
(Since $\psi_i$ was chosen to be real, there is no need to follow up with an ordinary impulse to correct the final phase.)
For a different choice of $\psi_i$, however, the same super impulse would generate a deformation that generically would not be equivalent to a rotation.

\section{Classical and semiclassical considerations}
\label{sec:classSemiclass}

In this section we first consider classical trajectories generated by the Hamiltonian $H_I$ (Eq.~\ref{eq:HI-3d}) in the fast time variable.
We show that these trajectories are equivalent to the Lagrangian trajectories of Sec.~\ref{sec:design3d}, and we relate them to the quantum propagator $K$ from $\psi_i$ to $\psi_f$, via the path integral formulation of quantum mechanics.
We then describe the evolution of a classical phase space density $\phi({\bf x},{\bf p},t)$ under a super impulse.
Finally, we establish a semiclassical connection between quantum and classical super impulses by considering an initial wavefunction that has the WKB form of a slowly (in space) varying amplitude modulated by a rapidly varying phase.

${\bf X}({\bf x}_i,\tau)$ obeys Newton's second law (Eqs.~\ref{eq:adef-3d}, \ref{eq:UI3d}),
\be
m\frac{\partial^2{\bf X}}{\partial\tau^2} = -\nabla U_2({\bf X},\tau) \quad .
\ee
If we define ${\bf P}({\bf x}_i,\tau) = m\,\partial{\bf X}/\partial\tau$, then ${\bf X}$ and ${\bf P}$ obey Hamilton's equations,
\be
\label{eq:hameq-main}
\frac{\partial{\bf X}}{\partial\tau} = \frac{\partial H_I}{\partial{\bf P}}
\quad ,\quad
\frac{\partial{\bf P}}{\partial\tau} = -\frac{\partial H_I}{\partial{\bf X}} \quad ,
\ee
with $H_I$ given by Eq.~\ref{eq:HI-3d}. 
In other words the Lagrangian trajectories of Sec.~\ref{sec:design3d} correspond to Hamiltonian trajectories of $H_I$.
These trajectories satisfy initial and final conditions
\begin{subequations}
\label{eq:XP}
\ba
({\bf X},{\bf P})_{\tau=0} &=& ({\bf x}_i,{\bf 0}) \\
({\bf X},{\bf P})_{\tau=T} &=& ({\bf x}_f({\bf x}_i),{\bf 0}) \quad .
\ea
\end{subequations}
Thus $H_I$ generates trajectories with the following property: if the system begins at rest, ${\bf P}_i={\bf 0}$, then it also ends at rest, ${\bf P}_f={\bf 0}$, and in between the motion in configuration space follows a straight line from ${\bf x}_i$ to ${\bf x}_f$ (Eq.~\ref{eq:Xdef-3d}).
The system's initial acceleration along this line is exactly compensated by later deceleration.
We will refer to this property by saying that the potential $U_2$ is {\it balanced.}
This property arises by careful design (see Eq.~\ref{eq:balance}) and is non-generic.
An arbitrarily chosen $U_2$ would generally be {\it unbalanced}: a vanishing momentum at $\tau=0$ would not imply a vanishing momentum at $\tau=T$.
We will return to this point in Sec.~\ref{sec:discussion}.

By the Hamilton-Jacobi equation (Eq.~\ref{eq:HJ3d}), the classical action along a trajectory obeying Eq.~\ref{eq:hameq-main} is given by $S({\bf X}({\bf x}_i,\tau),\tau)$~\cite{Littlejohn1992}, which vanishes at $t=T$ (Eq.~\ref{eq:S0T-3d}):
\begin{equation}
\label{eq:Sfinal}
S({\bf X}({\bf x}_i,T),T) = 0 \quad  \forall\, {\bf x}_i \quad .
\end{equation}
With these observations in mind, let us consider super impulses from the perspective of the path integral formulation of quantum mechanics.

Evolution under the time-dependent Schr\"odinger equation can be expressed in terms of a propagator $K({\bf x},t \vert {\bf x}^\prime,t^\prime)$, whose value is a sum over paths from ${\bf x}^\prime$ at time $t^\prime$ to ${\bf x}$ at a later time $t$~\cite{FeynmanHibbs_book}.
Each path contributes a phase $e^{iS/\hbar}$ given by the path's classical action.
In the semiclassical limit $\hbar\rightarrow 0$, the interference between these phases causes the propagator to be dominated by paths that correspond to classical trajectories.

For a super impulse, if we express the evolution from $\psi_i$ to $\psi_f$ in terms of a quantum propagator $K$, i.e.
\be
\psi_f({\bf x}) = \int d^D x_i \, K({\bf x},\epsilon T\vert{\bf x}_i,0) \, \psi_i({\bf x}_i) \quad ,
\ee
then Eq.~\ref{eq:psif} implies
\be
\label{eq:K}
K({\bf x},\epsilon T \vert {\bf x}_i,0) = \delta\Bigl( {\bf x} - {\bf x}_f({\bf x}_i) \Bigr) \, \left\vert \frac{\partial {\bf x}_f}{\partial {\bf x}_i} \right\vert^{1/2} \quad ,
\ee
which reveals that the propagator from ${\bf x}_i$ to ${\bf x}$ is determined entirely by a single path, namely the Hamiltonian trajectory satisfying ${\bf x}(0)={\bf x}_i$ and ${\bf p}(0) = {\bf 0}$.

The result that $K({\bf x},\epsilon T \vert {\bf x}_i,0)$ is determined by a single classical trajectory (for fixed ${\bf x}_i$ and as $\epsilon\rightarrow 0$) may seem suspicious.
Intuitively, we expect contributions from non-classical paths to vanish (due to interference) only in the semiclassical limit $\hbar\rightarrow 0$, whereas we have not assumed $\hbar$ to be small.
To explore this issue, let us write the Schr\"odinger equation \ref{eq:tdse} in the fast time variable $\tau=t/\epsilon$.
Multiplying both sides by $\epsilon^2$ gives, for $\tau\in[0,T]$,
\be
\label{eq:tdse-rescaled}
i(\epsilon\hbar) \frac{\partial\psi}{\partial\tau} = -\frac{(\epsilon\hbar)^2}{2m} \nabla^2 \psi  +  \epsilon^2 U_0({\bf x},\epsilon\tau) \psi  + U_2 ( {\bf x},\tau ) \psi \quad
\quad .
\ee
The second term on the right vanishes as $\epsilon\rightarrow 0$.
The remaining terms have the form of the time-dependent Schr\"odinger equation with Hamiltonian $H_I$ and an effective reduced Planck's constant:
\be
\label{eq:hbar_eff}
i\hbar_{\rm eff} \frac{\partial\psi}{\partial\tau} = -\frac{\hbar_{\rm eff}^2}{2m} \nabla^2\psi + U_2 ( {\bf x},\tau ) \psi \quad , \quad \hbar_{\rm eff} = \epsilon\hbar \quad .
\ee
This equation governs the evolution from $\psi_i$ at $\tau=0$ to $\psi_f$ at $\tau=T$.
We see that the impulsive limit ($\epsilon\rightarrow 0$ with $\hbar$ fixed, hence $\hbar_{\rm eff}\rightarrow 0$) acts as a proxy for the usual semiclassical limit ($\hbar\rightarrow 0$).
It is therefore not surprising that as $\epsilon\rightarrow 0$, phase interference suppresses the contributions to $K$ from non-classical paths.

Having related the quantum propagator $K({\bf x},\epsilon T \vert {\bf x}_i,0)$ to classical trajectories generated by the Hamiltonian $H_I$, let us now explicitly consider the evolution of a classical system under a super impulse.
Suppose we have designed a potential $U_2({\bf x},\tau)$ following the steps outlined in Sec.~\ref{subsec:satisfied}.
Let $\phi({\bf x},{\bf p},t)$ denote a classical phase space distribution that evolves under the Liouville equation
\be
\frac{\partial \phi}{\partial t} + \frac{\partial\phi}{\partial {\bf x}} \cdot \frac{\partial H}{\partial {\bf p}} - \frac{\partial\phi}{\partial {\bf p}} \cdot \frac{\partial H}{\partial {\bf x}} = 0 \quad ,
\ee
with $H$ given by Eq.~\ref{eq:H} and $k=2$.
Let $\phi_i$ and $\phi_f$ denote the distributions at $t=0$ and $\epsilon T$.
As shown in Appendix~\ref{sec:csuper}, a trajectory that starts at the phase point $({\bf x}_i,{\bf p}_i)$ at $t=0$ evolves to (see Eq.~\ref{eq:xfpf-class})
\be
\label{eq:xfpf}
({\bf x}_f,{\bf p}_f) = \Bigl( \mu({\bf x}_i) , L({\bf x}_i) \, {\bf p}_i  \Bigr)
\ee
at $t=\epsilon T$, in the limit $\epsilon\rightarrow 0$.
Here,
\be
\label{eq:Ldef}
L({\bf x}_i) = \frac{\partial{\bf p}_f}{\partial{\bf p}_i}({\bf x}_i,{\bf 0})
\ee
is a $D\times D$ matrix.
When ${\bf p}_i={\bf 0}$, Eq.~\ref{eq:xfpf} implies ${\bf p}_f={\bf 0}$.

Similarly defining
\be
\label{eq:Jdef}
J({\bf x}_i) = \frac{\partial{\bf x}_f}{\partial{\bf x}_i}({\bf x}_i,{\bf 0}) = \frac{\partial\mu({\bf x}_i)}{\partial{\bf x}_i} \quad,
\ee
and combining Eqs.~\ref{eq:xfpf} - \ref{eq:Jdef} with
\be
\label{eq:JTL-main}
J^T L = I
\ee
(derived in Appendix \ref{sec:csuper}), we obtain:
\begin{subequations}
\ba
\label{eq:deform-classical-a}
\phi_f({\bf x}_f,{\bf p}_f)
&=& \left\vert \frac{\partial({\bf p}_f,{\bf x}_f)}{\partial({\bf p}_i,{\bf x}_i)} \right\vert^{-1} \phi_i({\bf x}_i,{\bf p}_i) \\
\label{eq:deform-classical-b}
&=& \left\vert J({\bf x}_i) \right\vert^{-1} \left\vert L({\bf x}_i) \right\vert^{-1} \phi_i({\bf x}_i,{\bf p}_i) \\
\label{eq:deform-classical-c}
&=& \phi_i({\bf x}_i,{\bf p}_i) \quad ,
\ea
\end{subequations}
in agreement with Liouville's theorem.
[In going from Eq.~\ref{eq:deform-classical-a} to Eq.~\ref{eq:deform-classical-b} we have used $\vert\partial{\bf x}_f/\partial{\bf p}_i\vert=0$, which follows from Eq.~\ref{eq:xfpf}, and to get to Eq.~\ref{eq:deform-classical-c} we have used Eq.~\ref{eq:JTL-main}.]
Integrating both sides of Eq.~\ref{eq:deform-classical-b} over ${\bf p}_f$ gives
\be
\label{eq:rhof_rhoi_class}
\rho_f({\bf x}_f) = \left\vert J({\bf x}_i) \right\vert^{-1} \rho_f({\bf x}_i) \quad ,
\ee
which is Eq.~\ref{eq:rho_fi}.
(This result also follows directly from Eq.~\ref{eq:xfpf}.)
We conclude that quantum and classical super impulses with the same $U_2$ produce the same transport of probability distributions in coordinate space.

In the case of one dimension, Eq.~\ref{eq:xfpf} becomes
\be
\label{eq:xfpf-1d}
(x_f,p_f) = \left( \mu(x_i) , \frac{p_i}{\gamma(x_i)} \right) \quad ,
\ee
where $\gamma=\partial x_f/\partial x_i$ by Eq.~\ref{eq:JTL-main}.
Thus while $x_i$ is transformed under a generally nonlinear map, $p_i$ is rescaled linearly by a factor that guarantees Liouville's theorem is satisfied.

We can gain semiclassical insight into the effect of a quantum super impulse by assuming that $\psi_i({\bf x})$ has the WKB form of a slowly varying amplitude modulated by a rapidly oscillating phase.
In the vicinity of a point ${\bf x}_i^0$, we have a wavetrain $\psi_i({\bf x}) = A e^{i{\bf k}\cdot{\bf x}}$.
Expanding $\mu({\bf x})$ to first order around ${\bf x}_i^0$, Eq.~\ref{eq:psif} implies that the super impulse transforms this wavetrain as follows, aside from an overall phase:
\be
\label{eq:wavetrain}
\begin{split}
&\left[ \psi_i({\bf x}_i\approx {\bf x}_i^0) =  A e^{i{\bf k}\cdot{\bf x}_i} \right] \\
&\rightarrow
\left[ \psi_f({\bf x}_f\approx {\bf x}_f^0) = \frac{A}{\sqrt{\vert J\vert}} e^{i{\bf k} \cdot J^{-1} {\bf x}_f} 
= \frac{A}{\sqrt{\vert J\vert}} e^{i{\bf k} L^T \cdot {\bf x}_f} \right]
\quad ,
\end{split}
\ee
where ${\bf x}_f^0=\mu({\bf x}_i^0)$, $J$ and $L$ are evaluated at ${\bf x}_i^0$, and we have used ${\bf x}_f \approx {\bf x}_f^0 + J ({\bf x}_i-{\bf x}_i^0)$.
The super impulse displaces the wavetrain from ${\bf x}_i^0$ to ${\bf x}_f^0$ and linearly transforms the wavector from ${\bf k}$ to ${\bf k}^\prime = L{\bf k}$, adjusting the local amplitude accordingly.
In a semiclassical interpretation, the local momentum is tranformed from ${\bf p}_i=\hbar {\bf k}$ to $\hbar {\bf k}^\prime = L{\bf p}_i$.
Thus the local effect of the super impulse, on a wavefunction in WKB form, corresponds to the phase space map
\be
{\bf x}_i \rightarrow {\bf x}_f = \mu({\bf x}_i)
\quad,\quad
{\bf p}_i \rightarrow {\bf p}_f = L({\bf x}_i )\, {\bf p}_i \quad ,
\ee
in agreement with Eq.~\ref{eq:xfpf}.

Eq.~\ref{eq:wavetrain} becomes particularly transparent in the case of one degree of freedom:
\be
\begin{split}
&\left[ \psi_i(x_i\approx x_i^0) =  A e^{i k x_i} \right] \\
&\rightarrow
\left[ \psi_f(x_f\approx x_f^0) = \frac{A}{\sqrt{\gamma}} e^{ikx_f/\gamma} 
\right]
\end{split}
\ee
with $\gamma = \partial x_f/\partial x_i$ evaluated at $x_i^0$.
If $\gamma>1$ then the wavetrain is stretched by a factor $\gamma$ (see e.g.\ Fig.~\ref{fig:global-1d-deform} in the region $\vert x \vert <b$), and the resulting longer wavelength reduces the local momentum by the same factor, $p_f = p_i/\gamma$, in agreement with Eq.~\ref{eq:xfpf-1d}.

\section{Hybrid impulses}
\label{sec:hybrid}

If a super impulse is followed immediately by an ordinary impulse, as in Sec.~\ref{subsec:unsatisfied}, the outcome combines Eqs.~\ref{eq:paintPhase} and \ref{eq:deform}:
\be
\label{eq:phasePaint+deform}
\psi_f({\bf x}_f) = e^{i\Delta S({\bf x}_f)/\hbar} \psi_i({\bf x}_i) \left\vert \frac{\partial {\bf x}_f}{\partial {\bf x}_i} \right\vert^{-1/2} \quad .
\ee
The super impulse deforms the wavefunction and the ordinary impulse then paints a phase.
We now show that the same outcome can be achieved in one go, using a {\it hybrid} impulse
\be
\label{eq:H-hybrid}
H({\bf x},{\bf p},t) = \frac{{\bf p}^2}{2m} + U_0({\bf x},t) +  \frac{1}{\epsilon} U_1\left( {\bf x},\tau \right) + \frac{1}{\epsilon^2} U_2\left( {\bf x},\tau \right) \quad.
\ee
As in earlier sections, $U_1$ and $U_2$ vanish for $\tau\notin[0,T]$.

Given a map that satisfies Eq.~\ref{eq:assumption}, and a function $\Delta S({\bf x})$, we design a hybrid impulse as follows.
First, we construct $U_2$ using the recipe given by Eqs.~\ref{eq:FX-3d} - \ref{eq:UI3d}, which involves Lagrangian trajectories ${\bf x}(\tau)$ defined by Eq.~\ref{eq:Xdef-3d}.
Next, $U_1$ is given by
\be
\label{eq:U1-hybrid}
U_1({\bf x},\tau) = - \Delta S({\bf x}_f({\bf x},\tau)) \, \nu(\tau) \quad,
\ee
where $\int_0^T \nu(\tau)\,d\tau = 1$ and ${\bf x}_f({\bf x},\tau)$ is the final point (at $\tau=T$) along the Lagrangian trajectory that passes through ${\bf x}$ at time $\tau$.

The construction of $U_1$ and $U_2$ described in the previous paragraph gives a hybrid impulse that transforms an initial $\psi_i$ to a final $\psi_f$ given by Eq.~\ref{eq:phasePaint+deform}.
This claim is established by following steps that are nearly identical to those of Sec.~\ref{subsec:satisfied}, with only a minor modification as described at the end of Appendix \ref{sec:qsuper3d-app}.

For a given ${\bf x}_f$, Eq.~\ref{eq:U1-hybrid} can be rewritten as
\be
\label{eq:U1-hybrid-alt}
U_1({\bf x}(\tau),\tau) = -\Delta S({\bf x}_f) \, \nu(\tau) \quad,
\ee
where the left side is evaluated along the Lagrangian trajectory that ends at ${\bf x}(T)={\bf x}_f$.
Just as in the case of a super impulse (see Sec.~\ref{sec:classSemiclass}), the propagator $K$ for a hybrid impulse is determined by a single classical trajectory.
However, the presence of $U_1$ in Eq.~\ref{eq:H-hybrid} now contributes a term $-\int_0^T d\tau \, U_1({\bf x}(\tau),\tau)$ to the trajectory's action.
By Eq.~\ref{eq:U1-hybrid-alt} this term is equal to $\Delta S({\bf x}_f)$.
Hence Eq.~\ref{eq:K} becomes, for a hybrid impulse,
\be
\label{eq:K-hybrid}
K({\bf x},\epsilon T \vert {\bf x}_i,0) = e^{i\Delta S({\bf x})/\hbar} \delta\Bigl( {\bf x} - {\bf x}_f({\bf x}_i) \Bigr) \, \left\vert \frac{\partial {\bf x}_f}{\partial {\bf x}_i} \right\vert^{1/2} \quad ,
\ee
which is equivalent to Eq.~\ref{eq:phasePaint+deform}, and combines the expressions for propagators for ordinary and super impulses, Eqs.~\ref{eq:K-ordinary}, \ref{eq:K}.

Thus to deform a given wavefunction $\psi_i$ under a map $\mu$ that is not the gradient of a convex function, we can either follow the two-step procedure of Sec.~\ref{subsec:unsatisfied} -- a super impulse followed by an ordinary impulse -- or else we can apply a hybrid impulse as described above.

\section{Discussion}
\label{sec:discussion}

Sec.~\ref{sec:design3d} of this paper shows how to design a super impulse potential $U_2$ that suddenly deforms a quantum wavefunction $\psi_i$ under an invertible map $\mu$.
When the map is the gradient of a convex function, the super impulse is {\it global}: any $\psi_i$ can be deformed under $\mu$ using the same potential $U_2$.
When the map does not satisfy this criterion, the super impulse is {\it local}.
In that situation $U_2$ depends on $\psi_i$, and the super impulse must be followed by (or applied simultaneously with) an ordinary impulse.

These results offer a potentially useful tool for manipulating quantum systems, for instance to prepare them in desired states.
To assess this tool's feasibility, one must account for practical limitations related to the speed with which the potential $U_2$ can be varied, and the degree of experimental control over its shape and magnitude.
Within these limitations, can the desired deformation be achieved to an acceptably good approximation?
In particular, how small must $\epsilon$ be in order for Eq.~\ref{eq:deform} to be reliable?
We expect that if (1)  the term $\epsilon^2 U_0$ in Eq.~\ref{eq:tdse-rescaled} is negligible, and (2) the effective Planck constant $\hbar_{\rm eff} = \epsilon\hbar$ is sufficiently small that a semiclassical treatment accurately solves Eq.~\ref{eq:hbar_eff}, then Eq.~\ref{eq:deform} will accurately describe the post-impulse state of the wavefunction.
Numerical simulations will help to clarify this issue, and ultimately super impulses may be tested in the laboratory.
Cold atoms, which have been used to validate and implement quantum shortcuts to adiabaticity~\cite{Couvert08,Schaff10,Schaff11,Bason12,Rohringer15,Ness18,Du16,Zhou-NJP18,DengChenu18,DengDiao18}, provide a potential platform for such tests.

While the analysis of Secs.~\ref{sec:ordinary}-\ref{sec:design3d} was developed using the time-dependent Schr\"odinger equation, it applies equally well to evolution under the Gross-Pitaevskii equation
\be
i\hbar \, \partial_t \psi({\bf x},t) = \left[ H + g \left\vert \psi({\bf x},t) \right\vert^2 \right] \psi({\bf x},t) \quad ,
\ee
often used to model the evolution of Bose-Einstein condensates (BECs).
Here $H$ is a single-particle Hamiltonian given by Eq.~\ref{eq:H}, including the impulse $\epsilon^{-k}U_k$, and $g\vert\psi\vert^2\psi$ models particle-particle interactions at the mean-field level.
Because $g\vert\psi\vert^2\psi$ remains finite during the interval $[0,\epsilon T]$, its effect on the wavefunction's evolution during the impulse vanishes when $\epsilon\rightarrow 0$.
Thus the ordinary and super impulses derived in Secs.~\ref{sec:ordinary}-\ref{sec:design3d} for unitary evolution can also be applied to manipulate the evolution of BECs, within a mean-field approximation.

We have approached super impulses as a design problem: how do we construct a potential $U_2$ that deforms a wavefunction $\psi_i$ under a map $\mu$?
One can turn the question around to ask: given a potential $U_2({\bf x},\tau)$ in the interval $\tau\in[0,T]$, and an initial wavefunction $\psi_i$, what is the effect of the corresponding super impulse?

To address this question, let $S({\bf x},\tau)$ solve the Hamilton-Jacobi equation
\be
\label{eq:HJ-again}
\frac{\partial S}{\partial\tau} + \frac{(\nabla S)^2}{2m} + U_2 = 0
\ee
for the given $U_2({\bf x},\tau)$, with $S({\bf x},0)=0$.
Substituting Eq.~\ref{eq:ansatz3d}
into $i\hbar\,\partial_t\psi = H\psi$, with $H$ given by Eq.~\ref{eq:H}, and following the steps taken in Sec.~\ref{sec:design3d}, we again obtain Eq.~\ref{eq:theta+continuity3d} for $\rho$ and $\theta$.
The wavefunction $\psi_f({\bf x}) = \psi({\bf x},\epsilon T)$ is therefore given by
\be
\label{eq:psif-unbalanced}
\begin{split}
\psi_f({\bf x}_f)
&= \sqrt{ \rho_f({\bf x}_f) } \,  e^{i\theta_f({\bf x}_f)} e^{iS({\bf x}_f,T)/\epsilon\hbar} \\
&= \sqrt{ \rho_i({\bf x}_i) } \,  \left\vert \frac{\partial {\bf x}_f}{\partial {\bf x}_i} \right\vert^{-1/2} e^{i\theta_i({\bf x}_i)} e^{iS({\bf x}_f,T)/\epsilon\hbar} \\
&= \psi_i({\bf x}_i) \left\vert \frac{\partial {\bf x}_f}{\partial {\bf x}_i} \right\vert^{-1/2} e^{iS({\bf x}_f,T)/\epsilon\hbar}
\end{split}
\ee
where $\rho_{i,f}$ and $\theta_{i,f}$ are evaluated at $\tau=0$ and $T$, and ${\bf x}_f = \mu({\bf x}_i)$ (see Eqs.~\ref{eq:rho_fi-1d}, \ref{eq:theta_fi}).

In Sec.~\ref{sec:design3d}, $S$ was designed to vanish at $\tau=T$.
By contrast, in Eq.~\ref{eq:psif-unbalanced} $S$ is determined from $U_2$, through Eq.~\ref{eq:HJ-again}, and in general $S({\bf x},T)\ne 0$.
As a result, the phase of $\psi_f$ given by Eq.~\ref{eq:psif-unbalanced} is ill-behaved as $\epsilon\rightarrow\infty$.
This behavior traces back to the fact that an arbitrarily chosen potential $U_2$ is {\it unbalanced}:
for a classical trajectory $({\bf x}(\tau),{\bf p}(\tau))$ evolving under Eq.~\ref{eq:hameq-main}, and for a generic choice of $U_2$, ${\bf p}(0)={\bf 0}$ does not imply ${\bf p}(T)={\bf 0}$ (see Sec.~\ref{sec:classSemiclass}).
But ${\bf p}(\tau) = \nabla S({\bf x},\tau)$~\cite{Littlejohn1992}, hence a non-vanishing ${\bf p}(T)$ implies a non-vanishing, non-constant $S({\bf x}_f,T)$ in Eq.~\ref{eq:psif-unbalanced}.

Thus in order for a quantum super impulse to produce a well-behaved final wavefunction $\psi_f$, the potential $U_2$ must be balanced, in the sense introduced in Sec.~\ref{sec:classSemiclass}.
By design, the recipe provided in Sec.~\ref{subsec:satisfied} leads to balanced potentials.

Recall from Sec.~\ref{sec:classSemiclass} that, as $\epsilon\rightarrow 0$,
\be
\label{eq:K-limit}
K({\bf x},\epsilon T \vert {\bf x}_i,0) \rightarrow \delta\Bigl( {\bf x} - {\bf x}_f({\bf x}_i) \Bigr) \, \left\vert \frac{\partial {\bf x}_f}{\partial {\bf x}_i} \right\vert^{1/2} \quad .
\ee
This does not imply that $K\rightarrow 0$ for all ${\bf x} \ne {\bf x}_f({\bf x}_i)$.
Rather, in this limit $K$ oscillates ever more rapidly with ${\bf x}$, except at ${\bf x}={\bf x}_f$.
When $K$ is integrated with $\psi_i$ to give $\psi_f$, only the classical transition ${\bf x}_i\rightarrow {\bf x}_f$ contributes.
If $U_2({\bf x},\tau)$ is linear or quadratic in ${\bf x}$, and if $U_0=0$, then the propagator $K({\bf x},\epsilon T \vert {\bf x}_i, 0)$ can be solved analytically for any $\epsilon>0$, which may provide insight into how the limit in Eq.~\ref{eq:K-limit}, and by extension Eq.~\ref{eq:deform}, is approached.

Eq.~\ref{eq:theta+continuity3d}, which governs the wavefunction's dynamics during the interval $[0,\epsilon T]$, was derived by combining the {\it Ansatz} of Eq.~\ref{eq:ansatz3d} with the Schr\" odinger equation to obtain Eq.~\ref{eq:manyTerms}, and then taking the limit $\epsilon\rightarrow 0$.
If we instead define $\Sigma \equiv S + \epsilon\hbar\theta$ and separate the real and imaginary terms in Eq.~\ref{eq:manyTerms}, we obtain (for finite $\epsilon$):
\begin{subequations}
\label{eq:madelung}
\ba
\label{eq:madelung1}
0 &=& 
\dot\rho + \nabla\cdot\left(\frac{\nabla\Sigma}{m}\rho\right)  \\
\label{eq:madelung2}
0 &=&
\dot\Sigma + \frac{(\nabla\Sigma)^2}{2m} + \epsilon^2 U_0 + U_2 - \frac{(\epsilon\hbar)^2}{2m} \frac{\nabla^2\sqrt{\rho}}{\sqrt{\rho}} ~~~
\ea
\end{subequations}
Eq.~\ref{eq:madelung1} is the continuity equation under the flow field ${\bf v} + \epsilon\hbar\nabla\theta/m$, while Eq.~\ref{eq:madelung2} is the Hamiltonian-Jacobi equation for a particle moving in a potential given by the last three terms on the right.
Eq.~\ref{eq:madelung} is equivalent to the Madelung equations~\cite{Madelung26,Bialnicki-Birula92} and arises in the de Broglie-Bohm formulation of quantum mechanics~\cite{Bohm52}.
The last term in Eq.~\ref{eq:madelung2} is Bohm's {\it quantum potential}, with an effective Planck's constant $\hbar_{\rm eff} = \epsilon\hbar$.
Upon taking the limit $\epsilon\rightarrow 0$, both the background potential $U_0$ and the quantum potential drop out of Eq.~\ref{eq:madelung2}.
In this limit, Bohmian particle trajectories become the Lagrangian trajectories ${\bf X}({\bf x}_i,\tau)$ of Sec.~\ref{sec:design3d}.

Sanz {\it et al}~\cite{Sanz12} use the Madelung equations to study light flow through a Y-junction optical waveguide, which resembles the wavefunction cleaving of Sec.~\ref{subsec:splitting}.
The optical streamlines of Ref.~\cite{Sanz12} correspond to the Lagrangian trajectories ${\bf X}({\bf x}_i,\tau)$ of the present paper.
It will be interesting to elucidate the optical counterpart of the limit $\epsilon\rightarrow 0$, and to consider whether analogues of super impulses have potential applications in the field of optical waveguides.

Scaling properties often provide useful tools for studying non-adiabatic quantum dynamics.
Deffner {\it et al}~\cite{Deffner14} have developed a general strategy for designing shortcuts to adiabaticity for scale-invariant driving.
Modugno {\it et al}~\cite{Modugno18} combine the Madelung equations with an effective scaling approach~\cite{Guery-Odelin02} to obtain approximate solutions of the Gross-Pitaevskii equation for the free expansion of a BEC, and Huang {\it et al}~\cite{Huang21} adopt a similar strategy to design shortcuts to adiabaticity for harmonically trapped BECs.
Bernardo~\cite{Bernardo20} has proposed to accelerate quantum dynamics by rescaling the entire Hamiltonian by a time-dependent factor $f(\tau)$. 
It will be interesting to explore potential applications of these and related scaling-based methods to the context of super impulses.

The Wasserstein distance, Eq.~\ref{eq:wasserstein}, was shown by Aurell {\it et al}~\cite{Aurell-PRL11,Aurell-JSP12} to be related to entropy production in Langevin processes, and has been used by Nakazato and Ito~\cite{Nakazato-PRR21}, Van Vu and Saito~\cite{VanVu-PRX23}, and Chennakesavalu and Rotskoff~\cite{Chennakesavalu-PRL23} to develop a geometric understanding of optimal (minimally dissipative) protocols for driven nonequilibrium systems.
These results establish a fascinating connection between the fields of optimal transport~\cite{Villani03,Peyre19} and stochastic thermodynamics~\cite{Seifert2012,PelitiPigolotti_book2021}.
Separately, Deffner~\cite{Deffner-NJP17} has used the Wasserstein distance~\footnote{
In Ref.~\cite{Deffner-NJP17} the primary focus is on the Wasserstein-1 distance, whereas Eq.~\ref{eq:wasserstein} and Refs.~\cite{Aurell-JSP12,Nakazato-PRR21,Chennakesavalu-PRL23} use the Wasserstein-2 distance.
}
to obtain a quantum speed limit for the Wigner representation of quantum states.
It will be interesting to explore further the connection between optimal transport and rapidly driven quantum systems, and in particular to attempt to develop a geometric framework encompassing quantum speed limits, shortcuts to adiabaticity and super impulses.

It is natural to consider whether our results can be extended to include charged particles in magnetic fields.
Masuda and Rice~\cite{Masuda15} have shown how magnetic fields can be used to rotate quantum wavefunctions rapidly, and Setiawan {\it et al}~\cite{Setiawan-PTEP23} have shown how such fields can generate non-equilibrium steady states.
Replacing ${\bf p}$ in Eq.~\ref{eq:H} by ${\bf p} -e{\bf A}/c$ leads to a term
proportional to ${\bf p}\cdot{\bf A} + {\bf A}\cdot{\bf p}$ in the Hamiltonian $H$.
Terms of this form often appear in the counterdiabatic approach to shortcuts to adiabaticity~\cite{Muga10,Ibanez12,Torrontegui13,Jarzynski13,dCampo13,Deffner14,Jarzynski17,Patra17,Kolodrubetz17,Guery-Odelin19}, and are related to fast-forward potentials by gauge transformations~\cite{dCampo13,Deffner14,Kolodrubetz17,Guery-Odelin19}.
It may be fruitful to explore this connection in the context of quantum impulses.

Finally, Carolan {\it et al}~\cite{Carolan-PRA22} have studied counterdiabatic control of systems driven smoothly from an adiabatic to an impulsive regime, and then back to adiabatic, via the Kibble-Zurek mechanism~\cite{Kibble80,Zurek85}.
They find that it is energetically efficient to apply counterdiabatic control only during the impulsive regime, that is when it is most urgently needed.
It will be interesting to clarify how their results relate to those of the present paper.

\acknowledgments{}

This research was supported by the U.S. National Science Foundation under Grant No. 2127900.
The author gratefully acknowledges helpful and stimulating discussions and correspondence with Erik Aurell, Sebastian Deffner, Adolfo del Campo, Wade Hodson, Jorge Kurchan, Anatoli Polkovnikov, Grant Rotskoff, and Pratyush Tiwary.

\appendix

\section{Invertibility of ${\bf X}({\bf x}_i,\tau)$}
\label{sec:invertibleX}

Consider two points ${\bf x}_i^{(1)} \ne {\bf x}_i^{(2)}$, and define
\be
\delta{\bf x} = {\bf x}_i^{(2)} - {\bf x}_i^{(1)} \quad,\quad \delta{\bf X} = {\bf X}({\bf x}_i^{(2)},\tau) - {\bf X}({\bf x}_i^{(1)},\tau) \quad .
\ee
We then have
\be
\begin{split}
\delta{\bf x} &\cdot \delta{\bf X} = \sum_{k=1}^D \delta x_k \int_0^1 ds \, \frac{d}{ds} X_k\left( {\bf x}_i^{(1)} + s \delta{\bf x},\tau \right) \\
&= \sum_{k=1}^D \delta x_k \int_0^1 ds \, \sum_{j=1}^D \frac{\partial X_k}{\partial x_j} \left( {\bf x}_i^{(1)} + s \delta{\bf x},\tau \right) \, \delta x_j \\
&= \int_0^1 ds \,  \sum_{j,k} \frac{\partial^2 F}{\partial x_j \partial x_k} \left( {\bf x}_i^{(1)} + s \delta{\bf x},\tau \right) \, \delta x_j \delta x_k \quad .
\end{split}
\ee
Since $F({\bf x},\tau)$ is convex (which follows from the convexity of $\Phi$, Eq.~\ref{eq:Phiconvex}), the integrand on the last line above is strictly positive for all $s$, hence
$\delta{\bf x} \cdot \delta{\bf X} > 0$ and therefore
\be
{\bf X}({\bf x}_i^{(2)},\tau) \ne {\bf X}({\bf x}_i^{(1)},\tau)  .
\ee
Thus no two points ${\bf x}_i^{(1)} \ne {\bf x}_i^{(2)}$ produce the same output ${\bf X}$, i.e.\ ${\bf X}({\bf x}_i,\tau)$ is invertible with respect to ${\bf x}_i$.

In fact, since ${\bf X}({\bf x}_i,\tau) = \nabla F({\bf x}_i,\tau)$ for convex $F$, it follows that ${\bf x}_i({\bf X},\tau)=\nabla G$, where
\be
G({\bf X},\tau) = \sup_{{\bf x}_i\in\mathbb{R}^D} \left\{ {\bf X}\cdot{\bf x}_i - F({\bf x}_i,\tau) \right\}
\ee
is the Legendre-Fenchel transform of $F$.

\section{Derivations of Eqs.~\ref{eq:nablaPsi}, \ref{eq:HJ3d}, \ref{eq:theta+continuity3d}, \ref{eq:phasePaint+deform}}
\label{sec:qsuper3d-app}

Letting $A({\bf x},{\bf x}_i,\tau)$ denote the quantity in square brackets on the first line of Eq.~\ref{eq:Psidef}, we have
\ba
\nabla \Psi({\bf x},\tau) &=& \frac{1}{g(\tau)} \left( \frac{\partial A}{\partial{\bf x}} + \frac{\partial A}{\partial{\bf x}_i} \cdot \frac{\partial{\bf x}_i}{\partial{\bf x}} \right) \Bigg\vert_{{\bf x}_i = {\bf x}_i({\bf x},\tau)} \nonumber\\
&=& \frac{1}{g(\tau)} \left[ {\bf x}-{\bf x}_i + \left( -{\bf x} + \frac{\partial F}{\partial{\bf x}_i} \cdot \frac{\partial{\bf x}_i}{\partial{\bf x}} \right) \right] \nonumber\\
&=& \frac{1}{g(\tau)} ( {\bf x} - {\bf x}_i ) = {\bf x}_f - {\bf x}_i \quad ,
\ea
which establishes Eq.~\ref{eq:nablaPsi}.
Here we have used the fact that $\partial A/\partial{\bf x}_i = -{\bf x} + \partial F/\partial{\bf x}_i$ vanishes when ${\bf x}_i = {\bf x}_i({\bf x},\tau)$, as follows from Eq.~\ref{eq:Xdef-3d}.

We also have
\be
\begin{split}
\frac{\partial\Psi}{\partial\tau} &=-\frac{\dot g}{g} \Psi + \frac{1}{g} \left[ \left( -{\bf x} + \frac{\partial F}{\partial{\bf x}_i} \right) \cdot \frac{\partial {\bf x}_i}{\partial\tau} + \frac{\partial F}{\partial\tau} \right] \\
&= -\frac{\dot g}{g} \Psi + \frac{1}{g} \frac{\partial F}{\partial\tau} \\
&= -\frac{\dot g}{g} \Psi - \frac{1}{2} \frac{\dot g}{g} \vert {\bf x}_i\vert^2 + \frac{\dot g}{g} \Phi({\bf x}_i) \\
&= -\frac{\dot g}{2} \vert {\bf x}_f-{\bf x}_i \vert^2 \quad ,
\end{split}
\ee
which combines with Eqs.~\ref{eq:nablaPsi}, \ref{eq:UI3d}, \ref{eq:Sdef3d} and \ref{eq:HI-3d} to give Eq.~\ref{eq:HJ3d}:
\be
\label{eq:HJ3d-appendix}
\begin{split}
\frac{\partial S}{\partial\tau} &= m\ddot g\Psi - \frac{m}{2} \dot g^2 \vert {\bf x}_f-{\bf x}_i \vert^2 \\
&= - U_2 - \frac{1}{2m} \left\vert \nabla S \right\vert^2 = -H_I \left( {\bf x},\nabla S,\tau \right)
\end{split} \quad .
\ee

Substituting Eq.~\ref{eq:ansatz3d} into the Schr\"odinger equation $i\hbar\partial_t\psi=H\psi$, with $H$ given by Eq.~\ref{eq:H3d}, and dividing both sides by $\psi$, we get
\be
\label{eq:manyTerms}
\begin{split}
\frac{i\hbar}{\epsilon} &\left( \frac{\dot\rho}{2\rho} + i\dot\theta + \frac{i\dot S}{\epsilon\hbar} \right) = -\frac{\hbar^2}{2m} \Biggl[
\frac{\rho\nabla^2\rho - (\nabla\rho)^2}{2\rho^2} + i\nabla^2\theta  \\ 
&+ \frac{i\nabla^2S}{\epsilon\hbar} + \frac{(\nabla\rho)^2}{4\rho^2} - (\nabla\theta)^2
- \frac{(\nabla S)^2}{\epsilon^2\hbar^2} + \frac{i\nabla\rho\cdot\nabla\theta}{\rho} \\
&+ \frac{i\nabla\rho\cdot\nabla S}{\rho\epsilon\hbar} - 2 \frac{\nabla\theta\cdot\nabla S}{\epsilon\hbar} \Biggr] + U_0 + \frac{U_2}{\epsilon^2} \quad .
\end{split}
\ee
Collecting terms by powers of $\epsilon$ gives
\be
\begin{split}
0 &= \frac{1}{\epsilon^2} \cancelto{=0}{ \left[ \dot S + \frac{(\nabla S)^2}{2m} + U_2 \right] } \\
&+ \frac{\hbar}{\epsilon} \left( -\frac{i\dot\rho}{2\rho} + \dot\theta - \frac{i\nabla^2S}{2m} - \frac{i\nabla\rho\cdot\nabla S}{2m\rho} + \frac{\nabla\theta\cdot\nabla S}{m} \right) \\
&-\frac{\hbar^2}{2m} \frac{\nabla^2\sqrt{\rho}}{\sqrt{\rho}} 
-\frac{\hbar^2}{2m} \left[ i\nabla^2\theta - (\nabla\theta)^2 \right] + U_0
\end{split}
\ee
The $\epsilon^{-2}$ terms cancel by Eq.~\ref{eq:HJ3d} / \ref{eq:HJ3d-appendix}.
Multiplying both sides by $\epsilon$, separating the real and imaginary terms, and using $\nabla S = m{\bf v}$ (Eq.~\ref{eq:gradS}), leads to
\be
\begin{split}
0 &= \dot\theta + {\bf v}\cdot\nabla\theta + {\cal O}(\epsilon) \\
0 &= \dot\rho + (\nabla\cdot{\bf v)}\rho + {\bf v}\cdot\nabla\rho + {\cal O}(\epsilon)
\end{split}
\ee
which in the limit $\epsilon\rightarrow 0$ gives Eq.~\ref{eq:theta+continuity3d}.

For a hybrid impulse (Eq.~\ref{eq:H-hybrid}), the quantity $U_2/\epsilon^2$ in Eq.~\ref{eq:manyTerms} is replaced by $(U_1/\epsilon) + (U_2/\epsilon^2)$, leading to the following evolution equations for $\theta({\bf x},\tau)$ and $\rho({\bf x},\tau)$, in the limit $\epsilon\rightarrow 0$:
\be
\begin{split}
\dot\theta + {\bf v}\cdot\nabla\theta &= -U_1/\hbar \\
\dot\rho + \nabla\cdot({\bf v}\rho) &= 0
\end{split}
\ee
hence
\ba
\theta({\bf x}_f,T) &=& \theta({\bf x}_i,0) - \frac{1}{\hbar} \int_0^\tau d\tau \, U_1({\bf x}(\tau),\tau) \nonumber\\
&=& \theta({\bf x}_i,0) + \frac{1}{\hbar} \int_0^\tau d\tau \, \Delta S({\bf x}_f) \, \nu(\tau) \nonumber\\
&=& \theta({\bf x}_i,0) + \Delta S({\bf x}_f)/\hbar \\
\rho({\bf x}_f,T) &=& \rho({\bf x}_i,0) \left\vert \frac{\partial {\bf x}_f}{\partial {\bf x}_i} \right\vert^{-1}
\ea
where ${\bf x}(\tau)$ is the Lagrangian trajectory evolving from ${\bf x}_i$ to ${\bf x}_f$, and we have used Eq.~\ref{eq:U1-hybrid}.
These results combine with Eq.~\ref{eq:ansatz3d} to give Eq.~\ref{eq:phasePaint+deform}.

\section{Classical Super Impulses}
\label{sec:csuper}

Consider a classical system evolving under Hamilton's equations of motion,
\be
\label{eq:hameq}
\frac{d{\bf x}}{dt} = \frac{\bf p}{m}
\quad,\quad
\frac{d{\bf p}}{dt} = -\nabla U_0({\bf x},t) - \frac{1}{\epsilon^2} \nabla U_2\left({\bf x},\frac{t}{\epsilon}\right) \quad.
\ee
Let ${\bf x}_f({\bf x}_i,{\bf p}_i;\epsilon)$ and ${\bf p}_f({\bf x}_i,{\bf p}_i;\epsilon)$ denote the final phase point of a trajectory that evolves during the interval $t\in[0 , \epsilon T]$, from initial conditions $({\bf x}_i,{\bf p}_i)$,  for a given $\epsilon$.
For fixed $({\bf x}_i,{\bf p}_i)$, we wish to solve for $({\bf x}_f,{\bf p}_f)$ in the limit $\epsilon\rightarrow 0$.
These dynamics describe a trajectory evolving under a classical super impulse.

Rewriting Eq.~\ref{eq:hameq} in terms of the variables
\be
\tilde{\bf x} = {\bf x} \quad,\quad \tilde{\bf p} = \epsilon {\bf p} \quad,\quad \tau = \frac{t}{\epsilon} \quad ,
\ee
we obtain
\be
\label{eq:hameq-newvars}
\frac{d\tilde{\bf x}}{d\tau} = \frac{\tilde{\bf p}}{m}
\quad,\quad
\frac{d\tilde{\bf p}}{d\tau} = -\epsilon^2 \nabla U_0\left(\tilde{\bf x},\epsilon\tau\right) -\nabla U_2(\tilde{\bf x},\tau) \quad .
\ee
Let $\tilde{\bf x}_f(\tilde{\bf x}_i,\tilde{\bf p}_i;\epsilon)$ and $\tilde{\bf p}_f(\tilde{\bf x}_i,\tilde{\bf p}_i;\epsilon)$ denote final conditions (at $\tau=T$) as functions of initial conditions and $\epsilon$.
Note that Eq.~\ref{eq:hameq-newvars} implies 
\be
\frac{\partial\tilde{\bf x}_f}{\partial\epsilon}(\tilde{\bf x}_i,\tilde{\bf p}_i;0) = 0 \quad,\quad \frac{\partial\tilde{\bf p}_f}{\partial\epsilon}(\tilde{\bf x}_i,\tilde{\bf p}_i;0) = 0
\ee

When evaluated at $\epsilon= 0$, Eq.~\ref{eq:hameq-newvars} gives
\be
m\frac{d^2\tilde{\bf x}}{d\tau^2} = - \nabla U_2  = m{\bf a}(\tilde{\bf x},\tau)
\ee
(see Eqs.~\ref{eq:vaPsi-3d}, \ref{eq:UI3d}).
Thus, given a Hamiltonian trajectory $(\tilde{\bf x}(\tau),\tilde{\bf p}(\tau))$ evolving from initial conditions $(\tilde{\bf x}_i,\tilde{\bf p}_i)$ under Eq.~\ref{eq:hameq-newvars}, with $\epsilon=0$, the coordinates $\tilde{\bf x}(\tau)$ satisfy the same second-order differential equation as the Lagrangian trajectories of Sec.~\ref{sec:design3d} (see Eq.~\ref{eq:adef-3d}).
If we further set $\tilde{\bf p}_i={\bf 0}$ so that $d\tilde{\bf x}/d\tau={\bf 0}$ at $\tau=0$, then $\tilde{\bf x}(\tau)$ becomes identical with the Lagrangian trajectory ${\bf X}(\tilde{\bf x}_i,\tau)$, which begins and ends at rest (Eqs.~\ref{eq:vdef-3d}, \ref{eq:v0T3d}).
Thus,
\be
\label{eq:XPf}
\tilde{\bf x}_f(\tilde{\bf x}_i,{\bf 0};0) = \mu(\tilde{\bf x}_i)
\quad,\quad
\tilde{\bf p}_f(\tilde{\bf x}_i,{\bf 0};0) = {\bf 0} \quad .
\ee

Because the transformation from initial to final conditions is canonical, the $2d\times 2d$ matrix
\be
{\cal M}
\equiv \left(
\begin{array}{cc}
\partial\tilde{\bf x}_f/\partial\tilde{\bf x}_i & \partial\tilde{\bf x}_f/\partial\tilde{\bf p}_i \\
\partial\tilde{\bf p}_f/\partial\tilde{\bf x}_i & \partial\tilde{\bf p}_f/\partial\tilde{\bf p}_i
\end{array}
\right)
\ee
(for any $\tilde{\bf x}_i$, $\tilde{\bf p}_i$, $\epsilon$) satisfies~\cite{Goldstein80}
\be
\label{eq:symplectic}
{\cal M}^T \Omega {\cal M} = \Omega \quad,\quad \Omega\equiv \left(
\begin{array}{rr}
0 &I \\
-I & 0
\end{array}
\right)
\ee
where $0$ and $I$ are the $D\times D$ null and identity matrices.
Eq.~\ref{eq:XPf}b gives
\be
\frac{\partial\tilde{\bf p}_f}{\partial\tilde{\bf x}_i}(\tilde{\bf x}_i,{\bf 0};0) = 0 \quad.
\ee
If we further define
\ba
J(\tilde{\bf x}_i) &\equiv& \frac{\partial\tilde{\bf x}_f}{\partial\tilde{\bf x}_i}(\tilde{\bf x}_i,{\bf 0};0) \\
L(\tilde{\bf x}_i) &\equiv& \frac{\partial\tilde{\bf p}_f}{\partial\tilde{\bf p}_i}(\tilde{\bf x}_i,{\bf 0};0) \quad ,
\ea
then Eq.~\ref{eq:symplectic}, evaluated at $\tilde{\bf p}_i={\bf 0}$ and $\epsilon=0$, implies
\be
\label{eq:JTL}
J^T L = I \quad .
\ee

Returning to the original variables $({\bf x},{\bf p})$, we obtain
\ba
{\bf x}_f({\bf x}_i,{\bf p}_i;\epsilon) &=& \tilde{\bf x}_f({\bf x}_i,\epsilon {\bf p}_i;\epsilon) \nonumber \\
&=&
\tilde{\bf x}_f({\bf x}_i,{\bf 0};0) + \frac{\partial \tilde{\bf x}_f}{\partial \tilde{\bf p}_i}({\bf x}_i,{\bf 0};0) \epsilon {\bf p}_i + {\cal O}(\epsilon^2) \nonumber \\
&=&
\mu({\bf x}_i) + {\cal O}(\epsilon) \\
{\bf p}_f({\bf x}_i,{\bf p}_i;\epsilon) &=& \frac{1}{\epsilon}\tilde{\bf p}_f({\bf x}_i, \epsilon {\bf p}_i\epsilon) \nonumber\\
&=& \frac{1}{\epsilon} \left[ \tilde{\bf p}_f({\bf x}_i,{\bf 0};0) + \frac{\partial \tilde{\bf p}_f}{\partial \tilde{\bf p}_i}({\bf x}_i,{\bf 0};0) \epsilon {\bf p}_i + {\cal O}(\epsilon^2) \right] \nonumber \\
&=& L({\bf x}_i) \,  {\bf p}_i + {\cal O}(\epsilon) 
\ea
Taking $\epsilon\rightarrow 0$ gives
\be
\label{eq:xfpf-class}
{\bf x}_f = \mu({\bf x}_i) \quad,\quad {\bf p}_f = L({\bf x}_i) \, {\bf p}_i \quad . 
\ee
Thus, under the Hamiltonian dynamics generated by a super impulse, Eq.~\ref{eq:hameq}, the coordinates ${\bf x}$ transform under the map $\mu$, while the momenta ${\bf p}$ transform linearly.

We note that Eq.~\ref{eq:JTL} implies
\be
\left\vert \frac{{\bf x}_f}{{\bf x}_i} \right\vert \, \left\vert \frac{{\bf p}_f}{{\bf p}_i} \right\vert = 1 \quad ,
\ee
which is a statement of Liouville's theorem for the evolution from $({\bf x}_i,{\bf p}_i)$ to $({\bf x}_f,{\bf p}_f)$ given by Eq.~\ref{eq:xfpf-class}.

\bibliographystyle{apsrev4-2}
\input{impulseControl.bbl}

\end{document}

%% file: impulseControl.bbl
%